\documentclass[10pt,twocolumn,letterpaper]{article}

\usepackage{cvpr}              

\definecolor{cvprblue}{rgb}{0.21,0.49,0.74}

\usepackage{graphicx}
\usepackage{amsmath}
\usepackage{amssymb}
\usepackage{booktabs}
\usepackage[normalem]{ulem}
\useunder{\uline}{\ul}{}
\usepackage{multirow}
\usepackage{makecell}
\usepackage{algorithm}
\usepackage{algpseudocode}
\usepackage{xcolor}
\usepackage{colortbl}
\usepackage[symbol]{footmisc}
\usepackage{fancyvrb}

\usepackage[pagebackref,breaklinks,colorlinks,allcolors=cvprblue]{hyperref}
\usepackage[accsupp]{axessibility} 

\def\paperID{8616} 
\def\confName{CVPR}
\def\confYear{2025}

\title{PosterO: Structuring Layout Trees to Enable Language Models \\ in Generalized Content-Aware Layout Generation}

\author{HsiaoYuan Hsu and Yuxin Peng\thanks{}\\
Wangxuan Institute of Computer Technology, Peking University\\
{\tt\small kslh99@stu.pku.edu.cn, pengyuxin@pku.edu.cn}}

\begin{document}
\twocolumn[{
\renewcommand\twocolumn[1][]{#1}
\maketitle
\begin{center}
    \vspace{-2.7\baselineskip}
    \centering
    \captionsetup{type=figure}
    \includegraphics[width=0.98\linewidth]{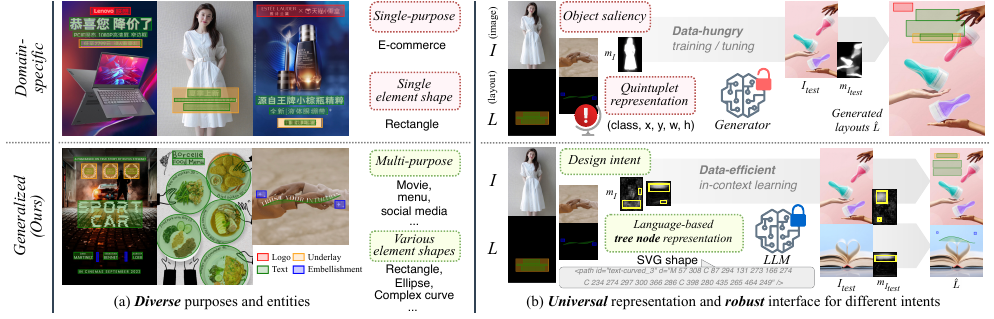}
    \vspace{-0.8\baselineskip}
    \caption{Illustration of the generalized settings in content-aware layout generation from both (a) data and (b) approach perspectives.}
    \label{fig:topic}
    \vspace{-0.45\baselineskip}
\end{center}
}]

\footnotetext[1]{Corresponding author.}

\begin{abstract}

\vspace{-0.7\baselineskip}

In poster design, content-aware layout generation is crucial for automatically arranging visual-textual elements on the given image.
With limited training data, existing work focused on image-centric enhancement.
However, this neglects the diversity of layouts and fails to cope with shape-variant elements or diverse design intents in \textbf{generalized} settings.
To this end, we proposed a layout-centric approach that leverages layout knowledge implicit in large language models (LLMs) to create \textbf{poster}s for \textbf{o}mnifarious purposes, hence the name \textbf{PosterO}.
Specifically, it structures layouts from datasets as trees in SVG language by \textbf{universal shape, design intent vectorization}, and \textbf{hierarchical node representation}.
Then, it applies LLMs during inference to predict new layout trees by in-context learning with \textbf{intent-aligned example selection}.
After layout trees are generated, we can seamlessly realize them into poster designs by editing the chat with LLMs.
Extensive experimental results have demonstrated that PosterO can generate visually appealing layouts for given images, achieving new state-of-the-art performance across various benchmarks.
To further explore PosterO's abilities under the generalized settings, we built \textbf{PStylish7}, the first dataset with multi-purpose posters and various-shaped elements, further offering a challenging test for advanced research.
%
Code and dataset are publicly available at 
\href{https://thekinsley.github.io/PosterO/}{https://thekinsley.github.io/PosterO/}.

\vspace{-1.2\baselineskip}

\end{abstract}

\section{Introduction}
\label{sec:intro}

\vspace{-0.7\baselineskip}

Content-aware layout generation makes the poster design process intelligent by automatically arranging visual-textual elements on given images.
As the demand for effective design grows rapidly, a mainstream platform surpasses 185 million monthly users \cite{canva_user}, and 90\% of survey participants take AI-empowered features as key considerations \cite{canva_ai}.
This technology is gaining more attention for its practical value in various applications \cite{wang-2024-ACMMM-prompt2poster,lin-2023-ACMMM-autoposter,weng-2024-CVPR-desigen,yang-2023-ACMMM-videothumb}, as shown in \cref{fig:topic}.

Unlike other generation tasks \cite{jia-2021-ICML-align,schuhmann-2022-NIPS-laion}, collecting image-layout pairs for the target task requires costly annotation and inpainting processes \cite{Min-2022-IJCAI-CGL,hsu-2023-CVPR-posterlayout}.
Therefore, existing work has incorporated object hints \cite{Cao-2022-ACMMM-ICVT,Min-2022-IJCAI-CGL}, prior information \cite{hsu-2023-CVPR-posterlayout,chai-2023-ACMMM-twostage}, retrieved references \cite{lin-2023-NIPS-layoutprompter,horita-2024-CVPR-RALF}, and augmented data \cite{seol-2024-ECCV-posterllama} to confront the challenge of training data scarcity.
However, because these are mainly image-centric enhancements without increasing the adaptability and diversity of layouts, they are prone to get trapped in the local solution space \cite{hsu-2023-icig-densitylayout}.
On the other hand, while leveraging related knowledge implicit in large language models (LLMs) \cite{roziere-2023-arXiv-codellama,ouyang-2022-NIPS-gpt35,touvron-2023-arXiv-llama2,dubey-2024-arXiv-llama3} showed the potential for layout-centric enhancements \cite{tang-2024-ICLR-layoutnuwa,feng-2024-layoutgpt}, existing work \cite{lin-2023-NIPS-layoutprompter,seol-2024-ECCV-posterllama} has suffered from the semantically impoverished layout representations.
The two primary issues are the monotonous rectangular shape of elements and the isolated visual constraints, which led to failures at capturing the complexity of real-world layouts and the intricate visual relationship between images and elements.

To pursue data efficiency and semantic richness required for advances in content-aware layout generation, we proposed \textbf{PosterO} to enable in-context learning (ICL) of LLMs by structuring \textbf{\textit{layout tree}} representations.
Specifically, it consists of \textit{three} procedures, including
(a) \textbf{layout tree construction} that jointly represents layout elements and design intents (\textit{i.e.}, available areas on images) as hierarchical nodes in SVG trees \cite{svgstandard},
(b) \textbf{layout tree generation} that dynamically selects intent-aligned learning examples for the test input to facilitate ICL, and
(c) \textbf{poster design realization} that integrates design materials into the generated layout trees to create posters in the subsequent chats with LLMs.

We conducted extensive experiments on public benchmarks \cite{Min-2022-IJCAI-CGL,hsu-2023-CVPR-posterlayout} and demonstrated that PosterO has achieved new \textit{state-of-the-art} performance.
It particularly shows stability in managing domain adaption problems \cite{xu-2023-CVPR-PDA} and spatial distribution shifts \cite{horita-2024-CVPR-RALF} that make the current approaches stumped.
In additional efforts, we enhanced the evaluation process through newly added intent-aware content metrics that are more credible than currently used saliency-aware ones.
We also demonstrated the adaptability of PosterO to different design intents and small-scale LLMs \cite{meta-2024-blog-llama32}, highlighting its applicability.
To further explore its potential under the \textbf{\textit{generalized}} settings of diverse poster purposes (\textit{e.g.}, commercial or education) and shape-variant elements (\textit{e.g.}, circles or curves), we built the first dataset, \textbf{PStylish7}, consisting of 152 few-shot learning samples and 100 test images, covering \textit{seven} representative poster purposes and \textit{eight} element types.
Serving as a challenging test, it is believed to facilitate advanced research in this field.

We summarize the contribution of this paper as follows:
\begin{itemize}
    \item A language-based content-aware layout representation, \textbf{layout tree}, uses hierarchical nodes to model both layout elements and design intents with arbitrary shapes.
    \item A layout-centric approach, \textbf{PosterO}, leverages implicit knowledge of LLMs by applying a few intent-aligned layout trees as examples to perform in-context learning.
    \item Extensive evaluations and newly added metrics verify the superior effectiveness of PosterO in confronting various adaptation problems that plague current approaches.
    \item A dataset for generalized content-aware layout generation, \textbf{PStylish7}, incorporates seven representative poster purposes and eight shape-variant element types.
\end{itemize}
As our best knowledge, this work is the first to consider the generalized settings of content-aware layout generation.
\section{Related Work}
\label{sec:related}

\subsection{Content-aware Layout Generation}

Apart from general layout tasks \cite{Li-2020-TPAMI-LayoutGAN,arroyo-2021-CVPR-VTN,Lee-2020-ECCV-NDN,Kikuchi-2021-ACMMM-latent,gupta-2021-ICCV-layouttransformer}, content-aware layout generation is a cross-modal task \cite{haoran-2023-CJE,dai-2023-CJE} integrating visual conditions, \textit{i.e.}, image canvases, which are inherent in various real-world applications \cite{wang-2024-ACMMM-prompt2poster,lin-2023-ACMMM-autoposter,weng-2024-CVPR-desigen,yang-2023-ACMMM-videothumb}, highlighting its practical utility.
With advances in deep generative models, data-centric approaches have become dominant in this field, categorized into GAN-based \cite{hsu-2023-CVPR-posterlayout,Min-2022-IJCAI-CGL,xu-2023-CVPR-PDA,hsu-2023-icig-densitylayout}, auto-regression-based \cite{Cao-2022-ACMMM-ICVT,cao-2024-IJMLC-ICVT+,horita-2024-CVPR-RALF,li-2024-CIKM-DET}, and diffusion model-based \cite{li-2023-CIKM-radm,chai-2023-ACMMM-twostage} approaches.

Entering into data construction of image-layout pairs, CGL-GAN \cite{Min-2022-IJCAI-CGL} leveraged an inpainting model \cite{Suvorov-2022-WACV-LaMa} to remove design elements from posters and a saliency object detection model \cite{Wang-2020-AAAI-PFPN} to give generative networks additional hints about image compositions.
Similar saliency-enhanced paradigm \cite{qin-2022-ECCV-ISNet,Li-2021-TMM-BASNet,qin-2019-CVPR-BASNet} has been adopted by most subsequent approaches.
Dealing with limited training data, DS-GAN \cite{hsu-2023-CVPR-posterlayout} introduced prior knowledge \cite{Guo-2021-CHI-Vinci,Li-2020-TVCG-attrlayoutgan} to arrange layout elements in a meticulously designed order, making the patterns existing in the data more efficiently mined.
RALF introduced retrieval augmentation, leveraging a composition-aware similarity evaluation model \cite{fu-2024-NIPS-dreamsim} to find the nearest neighbors for a given image, and incorporating their layout features \cite{Kikuchi-2021-ACMMM-latent} as additional inputs for the generative network.
Moreover, Chai et al. \cite{chai-2023-ACMMM-twostage} adapted a general layout diffusion model \cite{chai-2023-CVPR-layoutdm} for content-aware tasks by incorporating aesthetic constraints and a saliency-aware layout plausibility ranker.

As current approaches are engaged in image-centric enhancement yet seldom expand the diversity of layouts, the accessible solution space is narrowed, impeding their ability to address spatially varying design intents across different applications.
In light of this, we propose a layout-centric, intent-aware approach in this work.

\subsection{Language Model-based Layout Generation}
While numerical layout representations risk losing the semantic information of attributes \cite{tang-2024-ICLR-layoutnuwa}, current advances in large language models (LLMs) \cite{touvron-2023-arXiv-llama2,dubey-2024-arXiv-llama3,roziere-2023-arXiv-codellama} open up new possibilities with HTML-based \cite{lin-2023-NIPS-layoutprompter,peng-2024-ECCV-dreamstruct}, CSS-based \cite{chen-2024-EMNLP-textlap,feng-2024-layoutgpt}, and SVG-based \cite{seol-2024-ECCV-posterllama,tang-2024-ICLR-layoutnuwa} language representations as these approaches enable leverage of layout knowledge implicit in LLMs.
However, due to the challenge of guiding LLMs to perceive visual content, only a few approaches \cite{lin-2023-NIPS-layoutprompter,seol-2024-ECCV-posterllama} have supported content-aware layouts.

LayoutPrompter \cite{lin-2023-NIPS-layoutprompter} represented each layout element as a {\ttfamily <div>} in HTML.
While adopting a saliency paradigm, it only extracted the minimum bounding rectangle from the binary map as coarse content constraints and then retrieved layout examples for in-context learning.
In contrast, PosterLlama \cite{seol-2024-ECCV-posterllama} adopted a SVG-based representation \cite{tang-2024-ICLR-layoutnuwa} using {\ttfamily <rect>} shapes.
It then fine-tuned DINOv2 \cite{oquab-2023-arXiv-dinov2,zhu-2023-arxiv-minigpt} with CodeLlama \cite{roziere-2023-arXiv-codellama,hu-2021-arXiv-lora} enhanced by data augmentation \cite{zhang-2023-ICCV-ControlNet}.

\begin{figure*}[t]
  \centering
  \includegraphics[width=\textwidth] {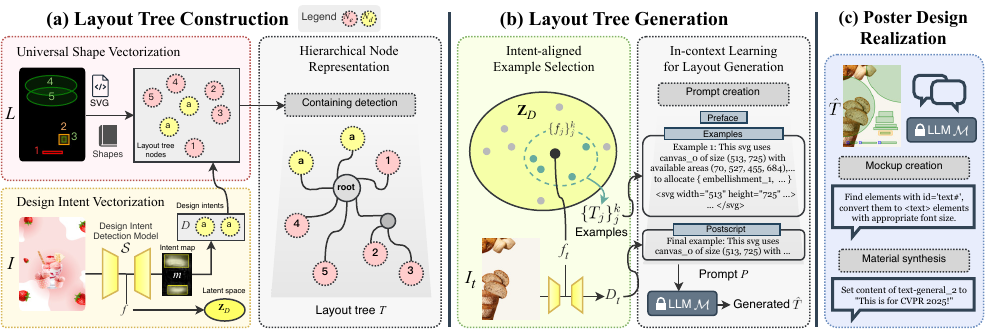}
    \vspace{-1.5em}
  \caption{An overview of PosterO. First, (a) takes image-layout pairs $(I, L)$ from datasets as input for data preparation, jointly modeling various-shaped elements $E$ and design intents $D$ towards layout trees $T$ and building up the latent space of $D$. Then, (b) takes test images $I_t$ as input and searches examples based on the predicted design intents $f_t$ to apply LLM $\mathcal{M}$ through in-context learning. After obtaining generated layout trees $\hat T$, (c) can continue the conversation with $\mathcal{M}$ to create poster designs seamlessly.}
  \label{fig:overview}
    \vspace{-1\baselineskip}  
\end{figure*}

As current approaches employed monotype element representations and isolated visual representations, they failed to exploit the rich semantics of languages nor explicitly capture the relationship between elements and images.
In light of this, we propose layout trees to represent various-shaped elements and visual hints jointly.
\section{The Proposed Approach: PosterO}
\label{sec:method}

Considering the limitations of image-centric and language model-based approaches, we propose PosterO.
It is layout-centric and enables large language models (LLMs) to generate visually appealing layouts for given images by structurally representing the content-aware layouts as \textit{trees} $T$ in SVG language.
An overview of PosterO is shown in \cref{fig:overview}.
In brief, it is composed of \textit{three} procedures, namely,
(a) \textbf{layout tree construction} for data preparation,
(b) \textbf{layout tree generation} applying an LLM $\mathcal{M}$ through $k$-shot in-context learning, and
(c) \textbf{poster design realization} through the subsequent chat with $\mathcal{M}$.

\subsection{Layout Tree Construction}
\label{subsec:ltc}
\subsubsection{Universal Shape Vectorization}

In content-aware layout generation, each data pair contains an image $I$ and a layout $L$.
Usually, $L$ is represented as a set of numerical tuples $\{e_i\}_i^n\!=\!\{(c_i, b_i)\}_i^n$, where $c_i$ and $b_i\!=\!(x, y, w, h)$ indicate the category and bounding box of the $i$-th element $e_i$ out of all $n$ elements.
As this is adequate only for rectangular elements, a more generalized representation becomes essential for the diverse shapes encountered in real-world posters, such as circles and curves.
In light of this, SVG (Scalable Vector Graphics) emerges as an ideal choice because of its ability to represent arbitrary shapes and, more importantly, its nature as a language to facilitate the layout knowledge implicit in LLMs.

\begin{figure}[t]
  \centering
  \includegraphics[width=\linewidth] {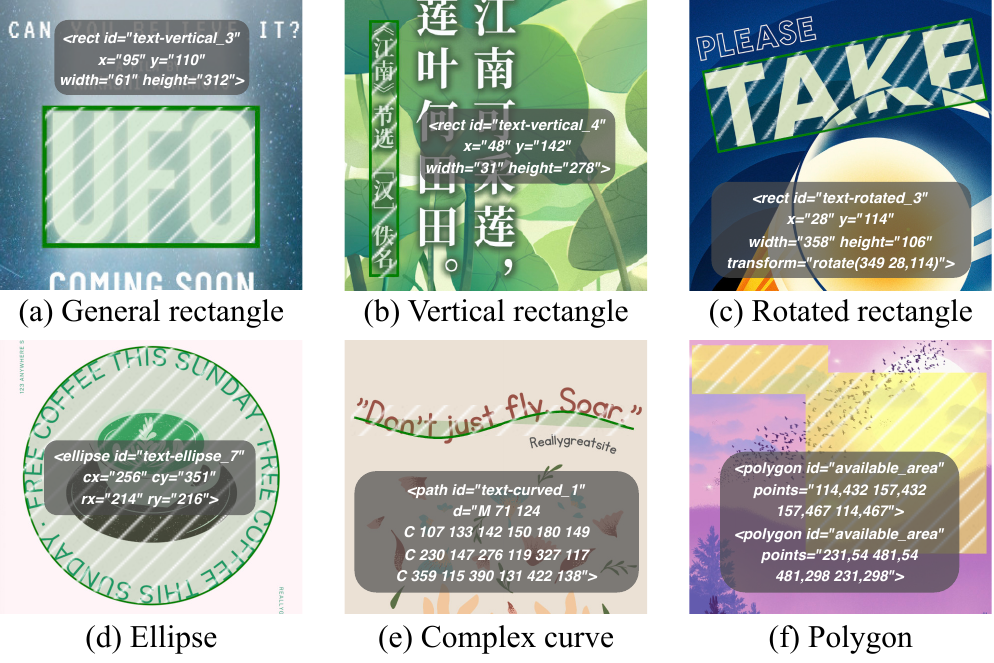}
  \vspace{-2em}
  \caption{Various shapes of elements in PStylish7 dataset.}
  \label{fig:univ-shape}
  \vspace{-1\baselineskip}
\end{figure}

Following the published SVG standards \cite{svgstandard}, we define five shapes to encompass a wide range of elements typically found in posters, as shown in \cref{fig:univ-shape}.
First, (a) \textbf{regular rectangle} is vectorized using {\ttfamily <rect>}. 
Also, its two variants, including
(b) \textbf{vertical rectangle} sensitive to the aspect ratio and
(c) \textbf{rotated rectangle} using the {\ttfamily rotate} function in the {\ttfamily transform} attribute to indicate the rotation degree.
When delving into more intricate shapes,
(d) \textbf{ellipse} is vectorized using {\ttfamily <ellipse>}, which can also handle the circle as a special case,
and (e) \textbf{complex curve} is vectorized using {\ttfamily <path>} approximated by multiple cubic Bézier curves.

\subsubsection{Design Intent Vectorization}
In addition to visual-textual elements, the design intents associated with the layout $L$ must also be vectorized.
Specifically, design intents $D$ are embodied by a set of areas within the image $I$, suitable or intended for placing the elements.
By incorporating these areas' representations $N_D$, the layout tree becomes content-aware and capable of conveying visual relationships between elements and images to LLMs, which cannot \!`see'\! the images.
As $D$ can be learned from existing data or customized case by case, this also provides a flexible interface for indicating various purposes.

By leveraging image-layout pairs from posters with a similar purpose, a design intent detection model $\mathcal{S}$ based on U-Net \cite{Ronneberger-2015-MICCAI-Unet} is trained in a semi-supervised manner.
Referring to \cite{hsu-2023-icig-densitylayout}, $\mathcal{S}$ takes the bitmap of layout elements as the supervision signal during the optimization, while the consistency loss is canceled to clearly segment out the areas.
With the pre-trained $\mathcal{S}$, PosterO can obtain design intents $D$ aligning the target purpose for any image $I$.
Specifically, after thresholding and morphological operations, we approximate the contour of the predicted intent maps $m$ and vectorize the obtained irregular areas using {\ttfamily <polygon>} in SVG language.
An example is shown in \cref{fig:univ-shape}(f).
Besides, the intermediate embeddings $f=\mathcal{S}_{\text{é}}(I)$, where $\mathcal{S}_{\text{é}}$ denotes the last block of the encoder, are also collected to serve as their representations in the latent space of design intents $\textbf{Z}_D$.

\subsubsection{Hierarchical Node Representation}
\label{subsec:hier}
As the nodes of layout elements ${N}_E$ and design intents ${N}_D$ are in place, the layout tree $T$ is constructed and encapsulated within an SVG container aware of the image resolution \!$(w_I\!, h_I)$, as:
\!{\ttfamily <svg\! width="$\!\{w_I\}\!$"\! height="$\!\{h_I\}\!$" x\!m\!l\!n\!s="\!h\!t\!t\!p\!:\!/\!/\!w\!w\!w\!.\!w\!3\!.\!o\!r\!g\!/\!2\!0\!0\!0\!/\!s\!v\!g\!"\!>}$\{{N}\!_D\}\{{N}\!_E\}${\ttfamily<\!/\!svg>}.
${N}_D$ is put ahead to ensure that ${N}_E$ is arranged considering their probabilistic dependencies in the autoregression, thus generating suitable content-aware layouts.
In addition, as the special elements that wrap around others, such as underlays, are usually used in poster design, a novel hierarchical representation explicitly modeling these enveloping properties is proposed.
Specifically, $N_E$ is sorted by prioritizing underlays with larger areas.
Then, given a node ${N}_a$ and the corresponding nodes ${N}_B\! =\! \{{N}_b \!\mid\! {N}_b \!\in\! \{N_E\}^n_{a+1} \wedge C({N}_a, $ ${N}_b)\}$, where the condition function $C(\cdot)$ is defined as:
\begin{equation}
    \begin{aligned}
        C(\cdot) = (\vert& {N}_a.x - {N}_b.x \vert \leq \epsilon) \wedge (\vert {N}_a.y - {N}_b.y \vert \leq \epsilon) \\
        \wedge&\ (\vert {N}_a.x + {N}_a.w - {N}_b.x - {N}_b.w \vert \leq \epsilon) \\
        \wedge&\ (\vert {N}_a.y + {N}_a.h - {N}_b.y - {N}_b.h \vert \leq \epsilon),
    \end{aligned}
\end{equation}
a subtree is constructed as {\ttfamily\!<svg\! x\!=\!"$\!\{{N}_a.x\}\!$"\! \!y\!=\!"$\!\{{N}_a.y\}\!$"> $\{{N}^\prime_a\}\{{N}^\prime_B\}$</svg>} and inserted at the position of ${N}_a$,
where the processed ${N}^\prime_i$ has the relative geometry attributes $({N}^\prime_i.x, {N}^\prime_i.y)\!=\!({N}_i.x\!-\!{N}_a.x,{N}_i.y\!-\!{N}_a.y)$.
All leaf nodes of $T$ are assigned unique identifiers {\ttfamily \!i\!d="$\{c_i\_i\}$"} to facilitate precise manipulation and querying.

\subsection{Layout Tree Generation}
\label{subsec:ltg}
\subsubsection{Intent-aligned Example Selection}
Upon findings from \cite{mann-2020-NIPS-GPT3}, LLMs are capable of in-context learning (ICL), requiring only a few examples to perform downstream tasks.
This significantly reduces the need for task-specific data and avoids the extensive costs or catastrophic forgetting \cite{kotha-2024-ICLR-undcf,luo-2023-arXiv-studycf} associated with fine-tuning.
To make the best use of layout knowledge implicit in LLMs during ICL, selecting examples that align with the design intent is crucial.
Therefore, given the test image $I_t$, its embedding $f_t$ is first extracted by the detection model $\mathcal{S}$;
subsequently, $k$ layout trees $\{T_j\}^k_j$ are selected as in-context examples, where their embeddings $\{f_j\}^k_j$ are the $k$ closest to $f_t$ in the latent space $\textbf{Z}_D$.

\subsubsection{In-context Learning for Layout Generation}
To provide clear structure and objectives in the prompt $P$ for the LLM $\mathcal{M}$, each intent-aligned example $T_j$ is preceded by a short sentence that informs its number $j$, the resolution of its corresponding image, design intents $D$, and identifiers of its nodes, as shown in \cref{fig:overview}(b).
The example field is then encapsulated within a preface and postscript, with the postscript delivering the request message of the test image $I_t$ through a short sentence mirroring that of the example.
Complete prompts $P$ are presented in supplementary material.
Finally, by inputting $P$ into $\mathcal{M}$ and interpreting the responses, PosterO completes the layout generation process.

\subsection{Poster Design Realization}
\label{subsec:pdr}
Benefiting from the essence of the generated layout tree $\hat T$ as graphics and the rigid {\ttfamily i\!d} assignment for elements, guiding $\mathcal{M}$ to transform $\hat T$ into a practical poster design is feasible and straightforward.
Specifically, the transformation contains \textit{mockup creation}, where shape elements are converted to {\ttfamily <text>} or {\ttfamily <image>} regarding their categories%
, and \textit{material synthesis}, where elements are filled with actual design contents, \textit{e.g.}, slogans and hyperlinks to images.
Notably, this transformation is achieved with zero-shot learning, yet $\mathcal{M}$ brings impressive results (in supplementary material) and even predicts suitable font sizes for texts, highlighting the value of design knowledge implicit in LLMs.
\section{Experiments}
\label{sec:experiment}
\subsection{Datasets}
To evaluate the proposed PosterO, two public e-commerce poster datasets are used, while their train+valid/annotated test/unannotated test splits are allocated as in \cite{horita-2024-CVPR-RALF} to ensure a fair comparison with existing work.
Specifically, \textbf{PKU PosterLayout} \cite{hsu-2023-CVPR-posterlayout} comprises 8,734/1,000/905 images with three layout element types, including text, logo, and underlay;
\textbf{CGL} \cite{Min-2022-IJCAI-CGL} comprises 54,546/6,002/1,000 images with four layout element types, including text, logo, underlay, and embellishment.
Furthermore, to explore PosterO's abilities under the generalized settings, we built the first multi-purpose poster dataset, \textbf{PStylish7}, containing layout elements of various shapes.
More details about PStylish7 are depicted in \cref{subsec:gclag} and supplementary material.

\begin{table*}[t]
    \centering

\begin{minipage}{0.7\textwidth}
    \resizebox{0.98\textwidth}{!}{%
    \begin{tabular}{c@{\hspace{10pt}}l|cccc|ccc|c} \toprule
&
Method &
  $Ove \downarrow$ &
  $Ali \downarrow$ &
  $Und_l \uparrow$ &
  $Und_s \uparrow$ &
  $Int \downarrow$ &
  $Sal \downarrow$ &
  $Rea \downarrow$ &
  $Avg \downarrow$ \\ \midrule
\rowcolor{Black!10}
\multicolumn{10}{c}{\textbf{Annotated Test Split}} \\
& \textcolor{gray!60}{\small \textit{Real data}}    & \textcolor{gray!60}{0.0010} & \textcolor{gray!60}{0.0038} & \textcolor{gray!60}{0.9942} & \textcolor{gray!60}{0.9903} & \textcolor{gray!60}{0.0289} & \textcolor{gray!60}{0.0368} & \textcolor{gray!60}{0.0109} & \textcolor{gray!60}{0.0138} \\
\multirow{5}{*}{{\rotatebox{90}{\small \textbf{Non-LLM}}}}
& CGL-GAN      & 0.0966 & 0.0035 & 0.7854 & 0.3570 & 0.1204 & 0.2774 & 0.0191 & 0.1964 \\
& DS-GAN       & {\ul 0.0261} & 0.0038 & {\ul 0.8350} & {\ul 0.5804} & \textbf{0.0677} & 0.3100 & 0.0199 & {\ul 0.1446} \\
& ICVT         & 0.2572 & 0.0405 & 0.5384 & 0.3932 & 0.7831 & 1.0836 & 0.0259 & 0.4655 \\
& LayoutDM\textsuperscript{\dag} & 0.1562 & \textbf{0.0018} & 0.6426 & 0.3873 & 0.5066 & {\ul 0.2429} & {\ul 0.0185} & 0.2709 \\
& RALF         & \textbf{0.0084} & {\ul 0.0028} & \textbf{0.9808} & \textbf{0.9201} & {\ul 0.1162} & \textbf{0.0479} & \textbf{0.0128} & \textbf{0.0410} \\
\midrule
\multirow{10}{*}{{\rotatebox{90}{\small \textbf{LLM-based}}}}
& \textit{\small Tuning-required}  &        &        &        &        &        &        &        &        \\
& PosterLlama\textsuperscript{\ddag}  & 0.0008 & 0.0006 & 0.9999 & 0.9982 & 0.3407 & 0.2619 & 0.0177 & 0.0891 \\
\cmidrule{2-10}
& \textit{\small Tuning-free}  &        &        &        &        &        &        &        &      \\
& LayoutPrompter$_{L2}$   & 0.0083 & 0.0030 & 0.3716 & 0.1633 & 0.3200 & 0.9433 & 0.0319 & 0.3959 \\
& LayoutPrompter$_{CL}$   & 0.0040 & 0.0030 & 0.4415 & 0.1937 & 0.3662 & 0.9035 & 0.0301 & 0.3817 \\
& LayoutPrompter$_{L3}$ & 0.0017 & 0.0028 & 0.4085 & 0.1613 & 0.5199 & 0.8567 & 0.0309 & 0.4060 \\
& PosterO$_{L2}$ (Ours) & \textbf{0.0004} & {\ul 0.0027} & 0.7597 & 0.5605 & 0.3679 & {\ul 0.1491} & {\ul 0.0188} & 0.1741 \\
& PosterO$_{CL}$   (Ours) & \textbf{0.0004} & 0.0028 & {\ul 0.9621} & {\ul 0.8606} & {\ul 0.2397} & 0.2295 & 0.0189 & {\ul 0.0955} \\
& PosterO$_{L3}$ (Ours) & {\ul 0.0006} & \textbf{0.0025} & \textbf{0.9888} & \textbf{0.9211} & \textbf{0.1080} & \textbf{0.1110} & \textbf{0.0171} & \textbf{0.0470} \\
\midrule
\rowcolor{Maroon!10}
\multicolumn{10}{c}{\textbf{Unannotated Test Split}} \\
\multirow{5}{*}{{\rotatebox{90}{\small \textbf{Non-LLM}}}}
& \textcolor{gray!60}{\small \textit{Train data}} & \textcolor{gray!60}{0.0012} & \textcolor{gray!60}{0.0027} & \textcolor{gray!60}{0.9969} & \textcolor{gray!60}{0.9900} & \textcolor{gray!60}{0.0000} & \textcolor{gray!60}{0.0000} & \textcolor{gray!60}{0.0108} & \textcolor{gray!60}{0.0040} \\
& CGL-GAN                            & 0.1010 & 0.0048 & 0.7326 & 0.2743 & {\ul 0.2275} & 0.7755 & 0.0327 & 0.3049 \\
& DS-GAN                             & {\ul 0.0248} & {\ul 0.0046} & {\ul 0.7859} & {\ul 0.4676} & \textbf{0.1659} & 0.5868 & 0.0320 & {\ul 0.2229} \\
& ICVT                               & 0.2786 & 0.0480 & 0.4939 & 0.3549 & 0.7947 & 1.3289 & 0.0347 & 0.5194 \\
& LayoutDM\textsuperscript{\dag}     & 0.1638 & \textbf{0.0029} & 0.5987 & 0.3695 & 0.5357 & {\ul 0.3172} & {\ul 0.0264} & 0.2968 \\
& RALF                               & \textbf{0.0175} & 0.0069 & \textbf{0.9548} & \textbf{0.8653} & 0.5986 & \textbf{0.1779} & \textbf{0.0231} & \textbf{0.1434} \\
\midrule
\multirow{10}{*}{{\rotatebox{90}{\small \textbf{LLM-based}}}} 
& \textit{\small Tuning-required}   &        &        &        &        &        &        &        &        \\
& PosterLlama\textsuperscript{\ddag} & 0.0006 & 0.0006 & 0.9986 & 0.9917 & 0.2667 & 0.3823 & 0.0285 & 0.0983 \\
\cmidrule{2-10}
& \textit{\small Tuning-free}  &        &        &        &        &        &        &        &        \\
& LayoutPrompter$_{L2}$   & 0.0095 & 0.0032 & 0.3401 & 0.1161 & 0.5349 & 1.2050 & 0.0408 & 0.4767 \\
& LayoutPrompter$_{CL}$   & 0.0032 & 0.0028 & 0.4052 & 0.1464 & 0.5148 & 1.1106 & 0.0395 & 0.4456 \\
& LayoutPrompter$_{L3}$ & 0.0010 & {\ul 0.0026} & 0.4054 & 0.1621 & 0.7005 & 1.0838 & 0.0412 & 0.4660 \\
& PosterO$_{L2}$ (Ours) & \textbf{0.0004} & 0.0031 & 0.7284 & 0.5188 & 0.4030 & {\ul 0.3351} & {\ul 0.0265} & 0.2173 \\
& PosterO$_{CL}$ (Ours) & {\ul 0.0005} & 0.0028 & {\ul 0.9595} & {\ul 0.8451} & {\ul 0.2796} & 0.3945 & 0.0273 & {\ul 0.1286} \\
& PosterO$_{L3}$ (Ours) & \textbf{0.0004} & \textbf{0.0025} & \textbf{0.9856} & \textbf{0.9241} & \textbf{0.1427} & \textbf{0.2131} & \textbf{0.0248} & \textbf{0.0677} \\
\bottomrule
    \end{tabular}
    }
    \vspace{-0.5em}
    \caption{Comparisons of quantitative results on PKU PosterLayout dataset.}
    \label{tab:res_pku}
\end{minipage}
\begin{minipage}{0.29\textwidth}
    \includegraphics[width=\textwidth] {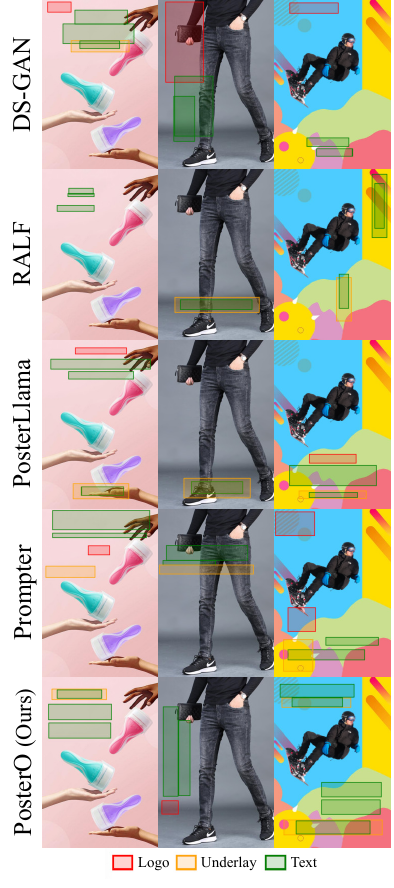}
    \vspace{-2.5em}
    \captionof{figure}{Comparisons of visualized results on the \textit{unannotated test split} of PKU PosterLayout dataset.}
    \label{fig:vis_pku}
\end{minipage}
  \vspace{-1.2\baselineskip}

\end{table*}

\subsection{Evaluation Metrics}
Following previous work \cite{hsu-2023-CVPR-posterlayout,Min-2022-IJCAI-CGL,horita-2024-CVPR-RALF}, we use graphic and content metrics.
\textbf{Graphic metrics} include overlay $Ove\!\!\downarrow$, alignment $Ali\!\!\downarrow$, and loose, strict underlay effectiveness $Und_l\!\!\uparrow$, $Und_s\!\!\uparrow$;
\textbf{Content metrics} include non-salient space utilization $Uti\!\uparrow$, salient space occlusion $Occ\!\!\downarrow$, and readability $Rea\!\!\downarrow$.
Moreover, we newly proposed two content metrics, design intent \textit{coverage} $Cov\!\!\uparrow$ and \textit{conflict} $Con\!\!\downarrow$, based on a design intent detection model $\mathcal{S}_{pc}$ that is comprehensively pre-trained on \cite{hsu-2023-CVPR-posterlayout} and \cite{Min-2022-IJCAI-CGL}.
Calculating $Cov\!\!\uparrow$ and $Con\!\!\downarrow$ is similar to $Uti\!\uparrow$ and $Occ\!\downarrow$ but using the predicted intent maps.
%
As reported in \cref{tab:sal_vs_den}, intent maps have \textbf{doubled} the matched portion with real layout data, reduced the unmatched portion by \textbf{34.7\%}, and required only \textbf{1.3\%} computational costs.
These findings have demonstrated that intent-aware metrics are more credible and efficient.
For a holistic understanding, we standardize intent-aware and saliency-aware metrics to $Int\!\downarrow$ and $Sal\!\downarrow$, as follows:
\begin{equation}
    \begin{aligned}
        Int &= \Sigma(\frac{\vert Cov_{\hat L} - Cov_{L}\vert}{1 - Cov_{L}}, \frac{\vert Con_{\hat L} - Con_{L}\vert}{Con_{L}}),\\
        Sal &= \Sigma(\frac{\vert Uti_{\hat L} - Uti_{L}\vert}{1 - Uti_{L}}, \frac{\vert Occ_{\hat L} - Occ_{L}\vert}{Occ_{L}}),\\
    \end{aligned}
\end{equation}
where $L$ and $\hat L$ are the layouts from the train split and generation.
Finally, we introduce an average metric $Avg\!\!\downarrow=\Sigma{(Ove, Ali, 1\!-\!Und_l, 1\!-\!Und_s, Int, Sal, Rea)}$.

\begin{table}[t]
\resizebox{\linewidth}{!}{%
\begin{tabular}{lcccc}
\toprule
\multirow{2}{*}{Prediction}        & \multicolumn{2}{c}{PKU}           & \multicolumn{2}{c}{CGL}           \\
\cmidrule(lr{0pt}){2-3} \cmidrule(lr{0pt}){4-5}
               & Match $\uparrow$ & Unmatch $\downarrow$ & Match $\uparrow$ & Unmatch $\downarrow$ \\ \midrule
Saliency map & 0.2238        & 0.1193          & 0.1984        & 0.1352          \\
Design intent map  & \textbf{0.4414}        & \textbf{0.0996}          & \textbf{0.4036}        & \textbf{0.0883}          \\ 
\bottomrule
\end{tabular}
}
\resizebox{\linewidth}{!}{%
\begin{tabular}{lccc} \toprule
Inference cost & Network architecture   & \#FLOPs(G)    & \#Params(M)     \\ \midrule
Saliency map & ISNet + BASNet            & 287.83        & 131.11          \\
Design intent map  & UNet (MiT-B1)       & \textbf{3.74} & \textbf{16.43}  \\
\bottomrule
\end{tabular}
}
\vspace{-0.2em}
\caption{Comparisons of the predicted saliency and design intent maps on the \textit{annotated test} splits of PKU PosterLayout and CGL datasets, showing that intent-aware metrics are more credible.}
    \label{tab:sal_vs_den}
    \vspace{-2\baselineskip}
\end{table}

\begin{table*}[t]
    \centering
    
\begin{minipage}{0.7\textwidth}
    \resizebox{0.98\textwidth}{!}{%
    \begin{tabular}{c@{\hspace{10pt}}l|cccc|ccc|c} \toprule
&
Method &
  $Ove \downarrow$ &
  $Ali \downarrow$ &
  $Und_l \uparrow$ &
  $Und_s \uparrow$ &
  $Int \downarrow$ &
  $Sal \downarrow$ &
  $Rea \downarrow$ &
  $Avg \downarrow$ \\
\midrule
\rowcolor{black!10}
\multicolumn{10}{c}{\textbf{Annotated Test Split}} \\
& \textcolor{gray!60}{\small \textit{Real data}} & \textcolor{gray!60}{0.0003} & \textcolor{gray!60}{0.0024} & \textcolor{gray!60}{0.9963} & \textcolor{gray!60}{0.9880} & \textcolor{gray!60}{0.0088} & \textcolor{gray!60}{0.0097} & \textcolor{gray!60}{0.0119} & \textcolor{gray!60}{0.0070} \\
\multirow{5}{*}{{\rotatebox{90}{\small \textbf{Non-LLM}}}}
& CGL-GAN      & 0.2291 & 0.0123 & 0.6466 & 0.2281 & 0.7088 & 0.4627 & 0.0213 & 0.3656 \\
& DS-GAN       & 0.0460 & {\ul 0.0022} & 0.9081 & 0.6308 & 0.3950 & 0.1539 & 0.0181 & 0.1537 \\
& ICVT         & 0.2453 & 0.0179 & 0.5150 & 0.3326 & 0.3632 & 0.5139 & 0.0211 & 0.3305 \\
& LayoutDM\textsuperscript{\dag} & {\ul 0.0184} & \textbf{0.0021} & {\ul 0.9216} & {\ul 0.8159} & {\ul 0.0411} & \textbf{0.0286} & \textbf{0.0137} & {\ul 0.0523} \\
& RALF         & \textbf{0.0042} & 0.0024 & \textbf{0.9912} & \textbf{0.9756} & \textbf{0.0366} & {\ul 0.0724} & {\ul 0.0180} & \textbf{0.0238} \\
\midrule
\multirow{10}{*}{{\rotatebox{90}{\small \textbf{LLM-based}}}}
& \textit{\small Tuning-required}  &        &        &        &        &        &        &        &        \\
& PosterLlama\textsuperscript{\ddag} & 0.0006 & 0.0006 & 0.9987 & 0.9890 & 0.1337 & 0.2559 & 0.0184 & 0.0602 \\
\cmidrule{2-10}
& \textit{\small Tuning-free}  &        &        &        &        &        &        &        &       \\
& LayoutPrompter$_{L2}$   & 0.0087 & 0.0033 & 0.3369 & 0.1455 & 0.3362 & 0.8234 & 0.0327 & 0.3888 \\
& LayoutPrompter$_{CL}$   & 0.0037 & 0.0031 & 0.3965 & 0.1616 & 0.3718 & 0.7799 & 0.0317 & 0.3760 \\
& LayoutPrompter$_{L3}$ & 0.0017 & 0.0030 & 0.3830 & 0.1740 & 0.5593 & 0.7957 & 0.0327 & 0.4051 \\
& PosterO$_{L2}$ (Ours)   & {\ul 0.0006} & {\ul 0.0023} & 0.7801 & 0.6150 & 0.2542 & {\ul 0.0467} & 0.0176 & 0.1323 \\
& PosterO$_{CL}$ (Ours)   & \textbf{0.0005} & {\ul 0.0023} & {\ul 0.9682} & {\ul 0.8868} & {\ul 0.1246} & 0.0700 & {\ul 0.0168} & {\ul 0.0513} \\
& PosterO$_{L3}$ (Ours) & {\ul 0.0006} & \textbf{0.0022} & \textbf{0.9842} & \textbf{0.9340} & \textbf{0.0388} & \textbf{0.0026} & \textbf{0.0161} & \textbf{0.0203} \\
\midrule
\rowcolor{Maroon!10}
\multicolumn{10}{c}{\textbf{Unannotated Test Split}} \\
& \textcolor{gray!60}{\small \textit{Train data}} & \textcolor{gray!60}{0.0003} & \textcolor{gray!60}{0.0027} & \textcolor{gray!60}{0.9945} & \textcolor{gray!60}{0.9858} & \textcolor{gray!60}{0.0000} & \textcolor{gray!60}{0.0000} & \textcolor{gray!60}{0.0118} & \textcolor{gray!60}{0.0049} \\
\multirow{5}{*}{{\rotatebox{90}{\small \textbf{Non-LLM}}}}
& CGL-GAN      & 0.2668 & 0.0316 & 0.6774 & 0.1656 & {\ul 0.9707} & 2.3973        & {\ul 0.0512} & 0.6964 \\
& DS-GAN       & 0.0991 & \textbf{0.0138} & {\ul 0.7566} & 0.2810 & \textbf{0.4607} & {\ul 2.2732} & 0.0526 & {\ul 0.5517} \\
& ICVT         & 0.2045 & 0.1010 & 0.4357 & 0.2599 & 1.2813 & 2.6512 & \textbf{0.0397} & 0.7974 \\
& LayoutDM\textsuperscript{\dag} & {\ul 0.0793} & 0.1822 & 0.6304 & {\ul 0.3853} & 1.5002 & 3.2906 & 0.0612 & 0.8711 \\
& RALF         & \textbf{0.0273} & {\ul 0.0189} & \textbf{0.9756} & \textbf{0.9315} & 1.0237 & \textbf{1.6798} & \textbf{0.0397} & \textbf{0.4118} \\
\midrule
\multirow{10}{*}{{\rotatebox{90}{\small \textbf{LLM-based}}}}
& \textit{\small Tuning-required}  &        &        &        &        &        &        &        &        \\
& PosterLlama\textsuperscript{\ddag}   & 0.0014 & 0.0007 & 0.9971 & 0.9771 & 0.3510 & 2.6170 & 0.0555 & 0.4359 \\
\cmidrule{2-10}
& \textit{\small Tuning-free}  &        &        &        &        &        &        &        &        \\
& LayoutPrompter$_{L2}$   & 0.0069 & 0.0103 & 0.2847 & 0.1126 & 1.0257 & 2.3702 & 0.0628 & 0.7255 \\
& LayoutPrompter$_{CL}$   & 0.0045 & {\ul 0.0024} & 0.3239 & 0.1341 & 1.5800 & 2.5380 & 0.0650 & 0.8188 \\
& LayoutPrompter$_{L3}$ & 0.0026 & \textbf{0.0016} & 0.2693 & 0.1142 & 1.6249 & 2.4128 & 0.0644 & 0.8175 \\
& PosterO$_{L2}$ (Ours)   & 0.0005 & 0.0033 & 0.6663 & 0.4770 & {\ul 0.3522} & \textbf{1.2365} & 0.0382 & 0.3553 \\
& PosterO$_{CL}$ (Ours)   & \textbf{0.0003} & 0.0037 & {\ul 0.9369} & {\ul 0.8071} & \textbf{0.2771} & 1.3223 & {\ul 0.0379} & {\ul 0.2710} \\
& PosterO$_{L3}$ (Ours) & {\ul 0.0004} & 0.0032 & \textbf{0.9816} & \textbf{0.9034} & 0.4328 & {\ul 1.2407} & \textbf{0.0365} & \textbf{0.2612} \\

\bottomrule
    \end{tabular}
}
    \caption{Comparisons of quantitative results on CGL dataset.}
    \label{tab:res_cgl}
\end{minipage}
\begin{minipage}{0.29\textwidth}
    \includegraphics[width=\textwidth] {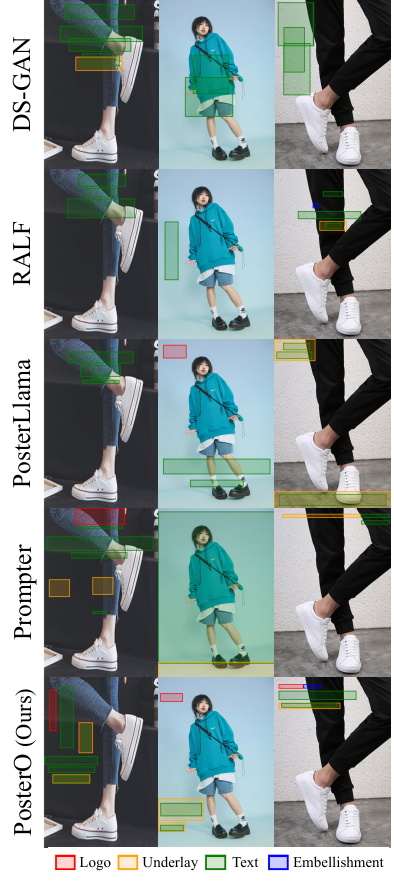}
    \vspace{-2em}
    \captionof{figure}{Comparisons of visualized results on the \textit{unannotated test split} on CGL dataset.}
    \label{fig:vis_cgl}
\end{minipage}
  \vspace{-1\baselineskip}

\end{table*}

\subsection{Implementation Details}
To ensure independent evaluations on \cite{hsu-2023-CVPR-posterlayout} and \cite{Min-2022-IJCAI-CGL}, we correspondingly train two detection models, $\mathcal{S}_{p}$ and $\mathcal{S}_{c}$, with mean squared error loss and a learning rate of $1e^{-6}$.
Each detection model is with a Transformer encoder tailored for semantic segmentation, MiT-B1 \cite{xie-2021-NIPS-segformer}.
For in-context learning (ICL), we implement $\mathcal{M}$ with three open-source LLMs, including Llama 2-7B $(L2)$ \cite{touvron-2023-arXiv-llama2}, CodeLlama-7B $(CL)$ \cite{roziere-2023-arXiv-codellama}, and Llama 3.1-8B $(L3)$ \cite{dubey-2024-arXiv-llama3}, with an unified sampling temperature of 0.7.
Considering the context lengths and remaining consistent with LayoutPrompter \cite{lin-2023-NIPS-layoutprompter}, the example size $k$ is 5, 10, and 10, respectively.
Still adhering to \cite{lin-2023-NIPS-layoutprompter}, a layout ranker is employed, while mIoU in the quality function measures the similarity between generated layouts and predicted design intents rather than real data, minimizing the risk of data leakage.

\subsection{Comparison with State-of-the-arts}
We select approaches with open-sourced implementation as baselines, including 
GAN-based CGL-GAN \cite{Min-2022-IJCAI-CGL}, DS-GAN \cite{hsu-2023-CVPR-posterlayout}, 
auto-regression-based ICVT \cite{Cao-2022-ACMMM-ICVT}, RALF \cite{horita-2024-CVPR-RALF}, 
diffusion model-based LayoutDM\footnote[2]{The extended version presented in \cite{horita-2024-CVPR-RALF}.} \cite{inoue-2023-CVPR-layoutdm}, 
and also LLM-based PosterLlama\footnote[3]{With the released weights tuned on \cite{hsu-2023-CVPR-posterlayout}, \cite{Min-2022-IJCAI-CGL}, and synthesized data.}\! \cite{seol-2024-ECCV-posterllama}, LayoutPrompter \cite{lin-2023-NIPS-layoutprompter}.
\vspace{-1.1\baselineskip}
\paragraph{Baseline comparison.}
\cref{tab:res_pku} reports the quantitative results on PKU PosterLayout.
The proposed PosterO shows the best overall performance across the test splits.
Particularly, when confronting the challenging unannotated split that poses domain adaptation problems \cite{xu-2023-CVPR-PDA}, it still achieves new \textit{state-of-the-art} (SOTA) results with outstanding stability.
Compared to the current non-LLM and LLM-based SOTA approaches, \textit{i.e.}, RALF and PosterLlama, ours significantly improves the average metric $Avg\!\downarrow$ by 52.8\% and 31.1\%.
Consistent results are obtained on CGL, as reported in \cref{tab:res_cgl}, ours improves $Avg\!\downarrow$ by 36.6\% and 40.1\%.

More precisely, PosterO shows impressive graphic metrics (2\textsuperscript{nd}\! column), closely resembling the distribution of real data, with only a slight deficit in the $Ali\!\downarrow$ metric on the unannotated split of CGL compared to LayoutPrompter.
As for content metrics (3\textsuperscript{rd}\! column), although PosterO uses only a small design intent detection model $\mathcal{S}$, it achieves comparable or even superior results to the baselines that utilize larger saliency detection models \cite{qin-2022-ECCV-ISNet,Li-2021-TMM-BASNet} or costly fine-tuned DINOv2 \cite{oquab-2023-arXiv-dinov2}.
Especially, PosterO is best at dealing with the \textit{spatial shifts in subject distributions} \cite{horita-2024-CVPR-RALF} present in the unannotated split of CGL, maintaining an overwhelming lead in content metrics.
\vspace{-1.1\baselineskip}
\paragraph{Backbone LLMs $\mathcal{M}$.}
Comparing the performance of tuning-free approaches across various LLMs is crucial for realizing how effectively the layout knowledge implicit in $\mathcal{M}$ has been leveraged.
As reported in \cref{tab:res_pku} and \cref{tab:res_cgl}, PosterO demonstrates overwhelming superiority over LayoutPrompter in all counterparts with the same $\mathcal{M}$.
More importantly, from PosterO$_{L2},_{CL}$, to PosterO$_{L3}$, it is evident that the proposed approach has experienced improvements as the capabilities of its based $\mathcal{M}$ increased.
This finding indicates that new knowledge, such as programming, introduced during model updates, has indeed been leveraged to generate layouts.
With the continuous development of LLMs, PosterO is anticipated to keep its rising potential.
\vspace{-1.3\baselineskip}
\paragraph{Visualized results.}
\cref{fig:vis_pku} and \cref{fig:vis_cgl} show the layouts generated by different approaches.
The results illustrate that PosterO generates high-quality layouts that the distribution of elements is generous and coordinated.
In contrast, the approaches based on training or tuning tend to place elements centrally and thus fail to confront images with compositions that are less common in the training data.
More visualized results are presented in supplementary material.

\begin{table}[t]
\centering
\resizebox{\linewidth}{!}{%
\begin{tabular}{l|cccccc}
\toprule
Method $\setminus$ $k$
               & 1      & 2      & 5      & 7      & 8      & 10     \\
\midrule
LayoutPrompter & 0.7711 & 0.6690 & 0.5353 & 0.4883 & 0.4919 & 0.4660 \\
PosterO (Ours) & 0.0678 & 0.0746 & 0.0732 & \uline{0.0663} & \textbf{0.0651} & 0.0677 \\
\bottomrule
\end{tabular}
}
\vspace{-0.5em}
    \caption{Comparisons of $Avg\!\downarrow$ over in-context example size $k$.}
    \label{tab:k}
  \vspace{-1\baselineskip}
\end{table}

\begin{figure}[t]
    \centering
    \includegraphics[width=\linewidth]{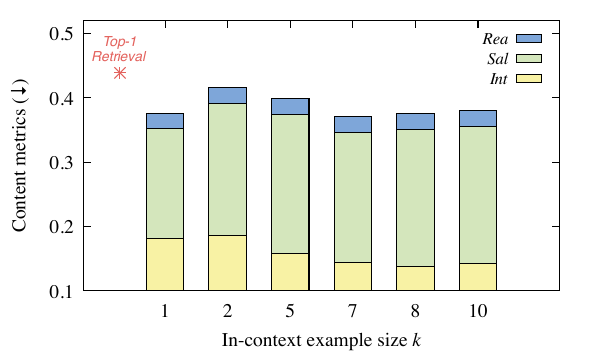}
    \vspace{-2em}
    \caption{Content metrics over in-context example size $k$.}
    \label{fig:k_icl}
  \vspace{-1\baselineskip}
\end{figure}
\vspace{-1.1\baselineskip}
\paragraph{In-context example size $k$.}
To explore the impact of example selection during ICL, we test a set of candidates $\{k\}\!=\!\{1,2,5,7,8,10\}$ on the unannotated split of \cite{hsu-2023-CVPR-posterlayout}, as reported in \cref{tab:k}.
Amazingly, PosterO consistently outperforms LayoutPrompter and all other baselines in \cref{tab:res_pku} across all settings of $k$.
As the optimal performance occurs at $k\!=\!7$ and $8$, it is proved that $\mathcal{M}$ has truly benefited from learning with the selected intent-aligned examples.
To further observe how well these examples help $\mathcal{M}$ understand visual contents, which is the focal point of content-aware layout generation, we plot the resultant content metrics over $k$ in \cref{fig:k_icl}.
It is noteworthy that directly adopting the most intent-aligned example (\textit{Top-1 Retrieval}) surpasses all baselines, indicating a successful alignment between intent and layout in the latent space $\mathbf{Z}_D$.
Nevertheless, we still observe improvements when involving in-context examples.

\subsection{Ablation Study}
To investigate our design choice in layout tree construction (\cref{subsec:ltc}) and learning example selection (\cref{subsec:ltg}), as well as to explore PosterO's adaptability to customized design intents and small-scale LLMs, we conducted the following ablation studies, mainly on the unannotated split of \cite{hsu-2023-CVPR-posterlayout}.

 \begin{table}[t]
\resizebox{\linewidth}{!}{%
\begin{tabular}{cc|cc|ccc|c}
\toprule
\textit{d} &
\textit{h} &
  $Und_l \uparrow$ &
  $Und_s \uparrow$ &
  $Int \downarrow$ &
  $Sal \downarrow$ &
  $Rea \downarrow$ &
  $Avg \downarrow$ \\ \midrule
$\surd$ & $\surd$ & \textbf{0.9856} & \textbf{0.9241} & \uline{0.1427} & \uline{0.2131} & 0.0248 & \textbf{0.0677} \\
\midrule
$\surd$ &         & 0.8609 & 0.6650 & \textbf{0.0981} & \textbf{0.1560} & \textbf{0.0240} & \uline{0.1078} \\
& $\surd$         & \uline{0.9497} & \uline{0.8836} & 0.9858 & 0.3991 & 0.0274 & 0.2260 \\
&                 & 0.7628 & 0.6413 & 0.8162 & 0.2729 & \uline{0.0246} & 0.2447 \\
\bottomrule
\end{tabular}
}
\vspace{-0.5em}
\caption{Ablation study on the (\textit{d}) design intent vectorization and (\textit{h}) hierarchical node representation in layout tree construction.}
\label{tab:ab_layouttree}
\end{table}
\vspace{-1.1\baselineskip}
\paragraph{Design choices for layout tree construction.}
We investigate the effectiveness of the design intent vectorization (\textit{d}) and hierarchical node representation (\textit{h}) during constructing layout trees, as reported in \cref{tab:ab_layouttree}.
Due to space constraints, $Ove\!\downarrow$ and $Ali\!\downarrow$ are omitted, considering their slight fluctuations.
As observed, the combination of \textit{d} and \textit{h} yields the best results.
Although removing \textit{h} appears to improve content metrics, this is primarily because the invalid empty underlays artificially inflate the utilization rate, as analyzed in supplementary material.
The significant decreases witnessed in underlay effectiveness, $Und_l$, $Und_s$, by up to 28\%, further underscore this issue.
In contrast, removing \textit{d} from both the examples and postscript makes LLMs blindly mimic the provided instances.
Here, however, shows a $Sal\!\downarrow$ surpasses some baselines, such as CGL-GAN, DS-GAN, and LayoutPrompter, while $Int\!\downarrow$ fails far behind, reflecting the robustness of the newly proposed metric $Int\!\downarrow$.
If both \textit{d} and \textit{h} are removed, the most severe degradation is observed, up to 261\%.
In this way, we demonstrate the necessity of these design choices.

\begin{figure}[t]
    \centering
    \includegraphics[width=0.99\linewidth]{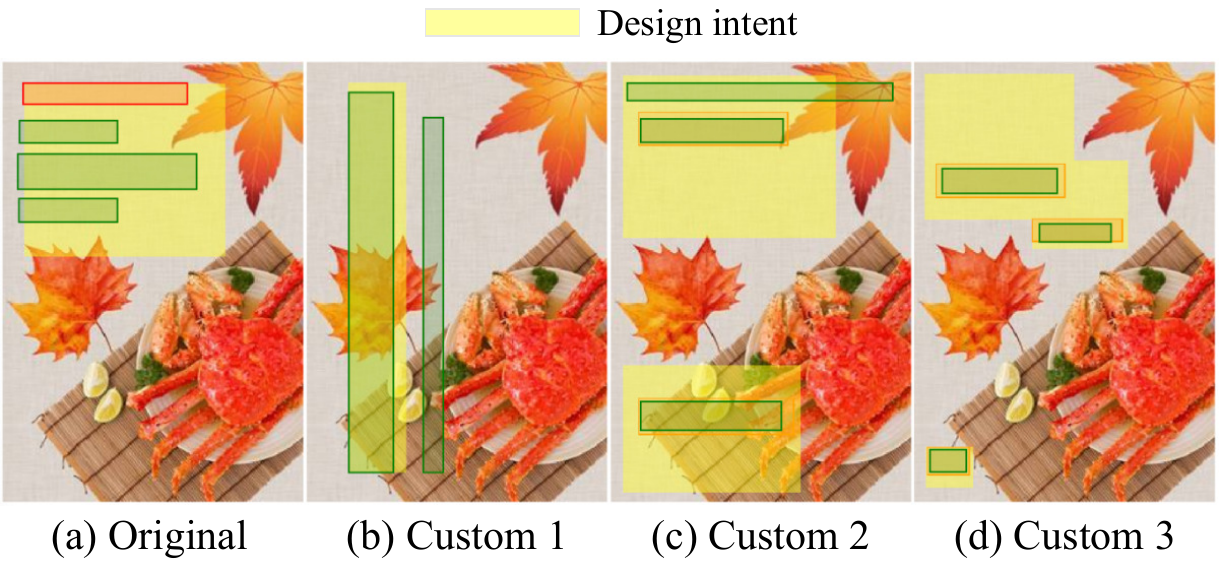}
    \vspace{-0.5em}
    \caption{PosterO's adaptability to different design intents $D$.}
    \label{fig:cus_d}
    \vspace{-0.5\baselineskip}
\end{figure}

 \begin{table}[t]
 \vspace{-0.5em}
\resizebox{\linewidth}{!}{%
\begin{tabular}{l|cc|ccc|c}
\toprule
Selection &
  $Und_l \uparrow$ &
  $Und_s \uparrow$ &
  $Int \downarrow$ &
  $Sal \downarrow$ &
  $Rea \downarrow$ &
  $Avg \downarrow$ \\ \midrule
$f$-aligned & 0.9856 & 0.9241 & \textbf{0.1427} & \textbf{0.2131} & \uline{0.0248} & \textbf{0.0677} \\
\midrule
$D$-aligned & \uline{0.9866} & \uline{0.9297} & \uline{0.2221} & \uline{0.2531} & \textbf{0.0236} & \uline{0.0836} \\
$E$-aligned & 0.9743 & 0.9026 & 0.3431 & 0.4177 & 0.0264 & 0.1304 \\
Random      & \textbf{0.9885} & \textbf{0.9302} & 0.3225 & 0.4688 & 0.0295 & 0.1293 \\
\bottomrule
\end{tabular}
}
\vspace{-0.5em}
\caption{Ablation study on different references for in-context example selection. ($f$: Embeddings of design intents, $D$: Bounding boxes of design intents, $E$: Sets of element types.)}
\vspace{-1.25\baselineskip}
\label{tab:ab_exsel}
\end{table}
\vspace{-1.1\baselineskip}
\paragraph{Design choices for in-context example selection.}
We have seen the remarkable performance of design intent embeddings $f$-aligned example selection, which relies on the learned latent space $\mathbf{Z}_D$.
To investigate alternative references, we have considered bounding boxes of design intents ($D$-aligned), sets of element types ($E$-aligned), and even with no specific alignment (Random).
As reported in \cref{tab:ab_exsel}, although all $D$-aligned, $E$-aligned, and random selections make declines in average metrics, the outcomes are still superior to most baseline approaches.
In particular, using $D$-aligned selection also achieves a new SOTA result, which supports its eligibility as an alternative enabling PosterO to deliver acceptable content-aware layouts when faced with roughly sketched $D$ from users.
\vspace{-1.1\baselineskip}
\paragraph{Adaption to customized design intent.}
Upon the discussions from the previous paragraph, we take a further step to verify PosterO's adaptability to customized design intents.
\cref{fig:cus_d} demonstrates the results of using the original $D$ predicted by the detection model $\mathcal{S}$ and three customized $D$ with different levels of complexity.
As the input $D$ varies, PosterO dynamically adjusts the spatial distribution of elements in the generated layouts while maintaining remarkable graphic metrics.
Notably, this capability is unique to PosterO, attributed to its intent-based paradigm.

\begin{table}[t]
\resizebox{\linewidth}{!}{%
\begin{tabular}{l|cc|ccc|c}
\toprule
Method &
  $Und_l \uparrow$ &
  $Und_s \uparrow$ &
  $Int \downarrow$ &
  $Sal \downarrow$ &
  $Rea \downarrow$ &
  $Avg \downarrow$ \\ \midrule
\rowcolor{Black!10}
\multicolumn{7}{c}{\small \textbf{Annotated Test Split of PKU PosterLayout}} \\
\textcolor{gray!60}{Ours$_{L3\text{-8B}}$}   & \textcolor{gray!60}{\textbf{0.9888}} & \textcolor{gray!60}{\textbf{0.9211}} & \textcolor{gray!60}{\textbf{0.1080}} & \textcolor{gray!60}{\uline{0.1110}} & \textcolor{gray!60}{\uline{0.0171}} & \textcolor{gray!60}{\textbf{0.0470}} \\
Ours$_{L3.2\text{-1B}}$ & 0.9159 & 0.6871 & 0.2633 & \textbf{0.1026} & 0.0174 & 0.1119 \\
Ours$_{L3.2\text{-3B}}$ & \uline{0.9662} & \uline{0.8595} & \uline{0.2166} & 0.1400 & \textbf{0.0168} & \uline{0.0787} \\ \midrule
\rowcolor{Maroon!10}
\multicolumn{7}{c}{\small \textbf{Unannotated Test Split of PKU PosterLayout}} \\
\textcolor{gray!60}{Ours$_{L3\text{-8B}}$}   & \textcolor{gray!60}{\textbf{0.9856}} & \textcolor{gray!60}{\textbf{0.9241}} & \textcolor{gray!60}{\textbf{0.1427}} & \textcolor{gray!60}{\textbf{0.2131}} & \textcolor{gray!60}{\textbf{0.0248}} & \textcolor{gray!60}{\textbf{0.0677}} \\
Ours$_{L3.2\text{-1B}}$ & 0.9008 & 0.6643 & 0.2702 & 0.2910 & 0.0276 & 0.1467 \\
Ours$_{L3.2\text{-3B}}$ & \uline{0.9731} & \uline{0.8724} & \uline{0.2276} & \uline{0.2479} & \uline{0.0249} & \uline{0.0940} \\ 
\midrule
\rowcolor{Black!10}
\multicolumn{7}{c}{\small \textbf{Annotated Test Split of CGL}} \\
\textcolor{gray!60}{Ours$_{L3\text{-8B}}$}   & \textcolor{gray!60}{\textbf{0.9842}} & \textcolor{gray!60}{\textbf{0.9340}} & \textcolor{gray!60}{\textbf{0.0388}} & \textcolor{gray!60}{\textbf{0.0026}} & \textcolor{gray!60}{\textbf{0.0161}} & \textcolor{gray!60}{\textbf{0.0203}} \\
Ours$_{L3.2\text{-1B}}$ & 0.9222 & 0.7724 & 0.1467 & 0.0463 & 0.0169 & 0.0741 \\
Ours$_{L3.2\text{-3B}}$ & \uline{0.9677} & \uline{0.8784} & \uline{0.0469} & \uline{0.0118} & \uline{0.0163} & \uline{0.0331} \\ \midrule
\rowcolor{Maroon!10}
\multicolumn{7}{c}{\small \textbf{Unannotated Test Split of CGL}} \\
\textcolor{gray!60}{Ours$_{L3\text{-8B}}$}   & \textcolor{gray!60}{\textbf{0.9816}} & \textcolor{gray!60}{\textbf{0.9034}} & \textcolor{gray!60}{0.4328} & \textcolor{gray!60}{1.2407} & \textcolor{gray!60}{\uline{0.0365}} & \textcolor{gray!60}{\uline{0.2612}} \\
Ours$_{L3.2\text{-1B}}$ & 0.8440 & 0.6409 & \textbf{0.2868} & \textbf{1.1101} & 0.0381 & 0.2791 \\
Ours$_{L3.2\text{-3B}}$ & \uline{0.9682} & \uline{0.8349} & \uline{0.3191} & \uline{1.1728} & \textbf{0.0359} & \textbf{0.2469} \\
\bottomrule
\end{tabular}%
}
\vspace{-0.3\baselineskip}
\caption{Ablation study on $\mathcal{M}$ accommodating small-scale LLMs.}
\label{tab:small_m}
\vspace{-1.2\baselineskip}
\end{table}
\vspace{-1.1\baselineskip}
\paragraph{Adaption to small-scale backbone LLM $\mathcal{M}$.}
To address resource-limited environments and to accelerate the layout generation process, we investigate PosterO's adaptability to small-scale LLMs \cite{meta-2024-blog-llama32}, including Llama3.2-1B $(L3.2\text{-1B})$ and Llama3.2-3B $(L3.2\text{-3B})$.
As reported in \cref{tab:small_m}, PosterO maintains very competitive and or even better results with the smaller $\mathcal{M}$, showcasing its potential for more flexible and efficient applications.

\subsection{Generalized Content-aware Layout Generation}
\label{subsec:gclag}
To verify PosterO's potential
in generalized content-aware layout generation, which is presented for the first time in this work, we built a new dataset, \textbf{PStylish7}.
It comprises 152 few-shot learning samples and 100 test images, covering seven poster categories. Each serves as a representative prototype for a segment in the spectrum of poster design purposes, encompassing artwork exhibition (\textit{Paint}), cultural education (\textit{Poem}), public safety (\textit{Metro}), entertainment marketing (\textit{Movie}), merchandising display (\textit{Menu}), public advocacy (\textit{Animal}), and social-media interaction (\textit{Instagram}).
Besides, it covers eight element types, including all those in the existing datasets and four text variants depicted in \cref{subsec:ltc}.
Considering feasibility, we employ five metrics, including $Ove\!\!\downarrow$, ($Cov\!\!\uparrow$, $Con\!\!\downarrow$), standardized as $Int\!\downarrow$, and ($Uti\!\uparrow$, $Occ\!\downarrow$), standardized as $Sal\!\downarrow$.
More details about PStylish7 are given in supplementary material.

\begin{table}[t]
\resizebox{\linewidth}{!}{%
\begin{tabular}{r|ccccccc}
\toprule
{\small Ours$_{CL}$} & \textit{Paint} & \textit{Poem}  & \textit{Metro} & \textit{Movie} & \textit{Menu}  & \textit{Animal} & \textit{Insta.} \\
\midrule
$Ove \downarrow$ & 0.0047 & 0.0034 & \textbf{0.0618} & 0.0106 & 0.0083 & \textbf{0.0032} & 0.0010 \\
$Int \downarrow$ & 0.7690 & \textbf{0.2198} & \textbf{0.6987} & \textbf{0.1216} & 0.2359 & \textbf{0.3020} & \textbf{0.2674} \\
$Sal \downarrow$ & 3.8838 & 0.8484 & 1.3604 & 0.3275 & 0.7191 & \textbf{2.6280} & \textbf{0.6652} \\
\midrule
$Avg \downarrow$ & 1.5525 & 0.3572 & 0.7070 & \textbf{0.1532} & 0.3211 & \textbf{0.9777} & \textbf{0.3112} \\
\bottomrule
\toprule
{\small Ours$_{L3}$} & \textit{Paint} & \textit{Poem}  & \textit{Metro} & \textit{Movie} & \textit{Menu}  & \textit{Animal} & \textit{Insta.} \\
\midrule
$Ove \downarrow$ & \textbf{0.0032} & \textbf{0.0028} & 0.0777 & \textbf{0.0004} & \textbf{0.0056} & 0.1495 & \textbf{0.0004} \\
$Int \downarrow$ & \textbf{0.6219} & 0.6274 & 0.7478 & 0.3375 & \textbf{0.1620} & 0.5383 & 1.1431 \\
$Sal \downarrow$ & \textbf{2.4512} & \textbf{0.1600} & \textbf{1.1111} & \textbf{0.1347} & \textbf{0.5790} & 2.9284 & 1.0810 \\
\midrule
$Avg \downarrow$ & \textbf{1.0254} & \textbf{0.2634} & \textbf{0.6455} & 0.1575 & \textbf{0.2489} & 1.2054 & 0.7415 \\
\bottomrule
\end{tabular}%
}
\vspace{-0.5\baselineskip}
\caption{Quantitative results on PStylish7 dataset.}
\vspace{-0.5\baselineskip}
\label{tab:pstylish}
\end{table}

\begin{figure}
    \centering
    \includegraphics[width=\linewidth]{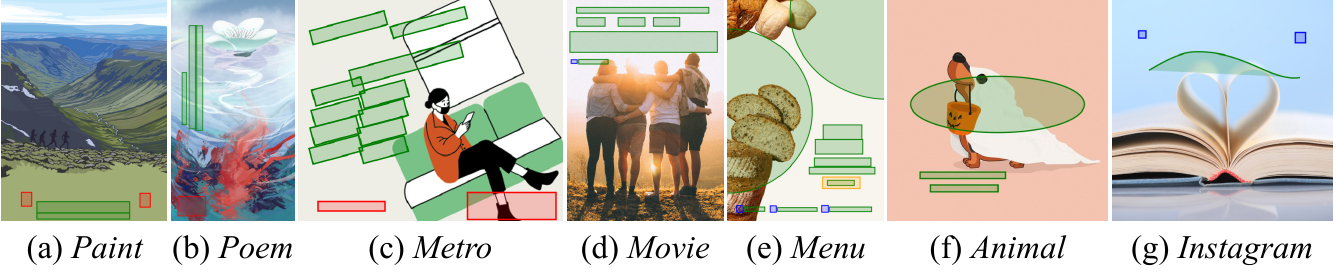}
    \vspace{-2em}
    \caption{Visualized results on PStylish7 dataset.}
    \label{fig:vis_ps}
\vspace{-1.25\baselineskip}
\end{figure}

The quantitative results are reported in \cref{tab:pstylish}, and the visualized results are presented in \cref{fig:vis_ps}.
Here, the detection model $\mathcal{S}_{pc}$ is employed to obtain design intents $D$ and their features $f\!$.
While PosterO$_{L3}$ outperforms PosterO$_{CL}$ in most categories, as the first attempt in this field, there is still plenty of room for improvement.
Overall, the proposed PStylish7 is significantly more challenging than the existing datasets, attributed to its broader and sparser data distribution.
This requires approaches with good generalization capabilities, mirroring real-world scenarios, and can stimulate more advanced work in the future.


\section{Conclusion and Discussion}
\label{sec:conclusion}
This work presented a tuning-free language model-based approach for content-aware layout generation, named \textbf{PosterO}.
Through extensive experiments, we demonstrated that PosterO achieves new state-of-the-art results on various benchmarks, mainly attributed to the structural representation, \textit{layout trees}.
Moreover, we delved into the task under \textit{generalized} settings and built the first available dataset, \textbf{PStylish7}, which covers seven poster purposes and eight element types.
Employing PStylish7, we verified PosterO's potential to solve complicated tasks in real-world scenarios, underscoring its practicality.
The future work mainly lies in two aspects.
First, enhancing the generalized capabilities of layout generation approaches is highly valuable, as PosterO aims.
With the challenging test offered by the proposed PStylish7 dataset, it is increasingly attainable.
Second, integrating feedback mechanisms is envisioned to facilitate self-optimization and allow the incorporation of user suggestions, forming a more interactive, user-centric process.

\section*{Acknowledgments}
This work was supported by the National Natural Science Foundation of China (61925201, 62132001, 62432001) and Beijing Natural Science Foundation (L247006).

{
    \small
    \bibliographystyle{ieeenat_fullname}
    \bibliography{main}
}

\clearpage
\maketitlesupplementary

\renewcommand\thesection{\Alph{section}}
\renewcommand\thetable{\Alph{table}}
\renewcommand\thefigure{\Alph{figure}}
\setcounter{page}{1}
\setcounter{section}{0}
\setcounter{table}{0}
\setcounter{figure}{0}
This supplementary material provides additional information about the proposed approach, \textbf{PosterO}, and the new dataset, \textbf{PStylish7}.
\cref{sec:supp_prompt} presents examples of layout trees $T$ and prompts $P$,
\cref{sec:supp_result} reports additional experimental results, and
\cref{sec:supp_ps7} provides more details and statistics about the PStylish7 dataset.
\vspace{-1\baselineskip}
\section{Examples of Layout Trees and Prompts}
\label{sec:supp_prompt}

\begin{figure}[b]
    \centering
    \vspace{-1.5\baselineskip}
\includegraphics[width=\linewidth]{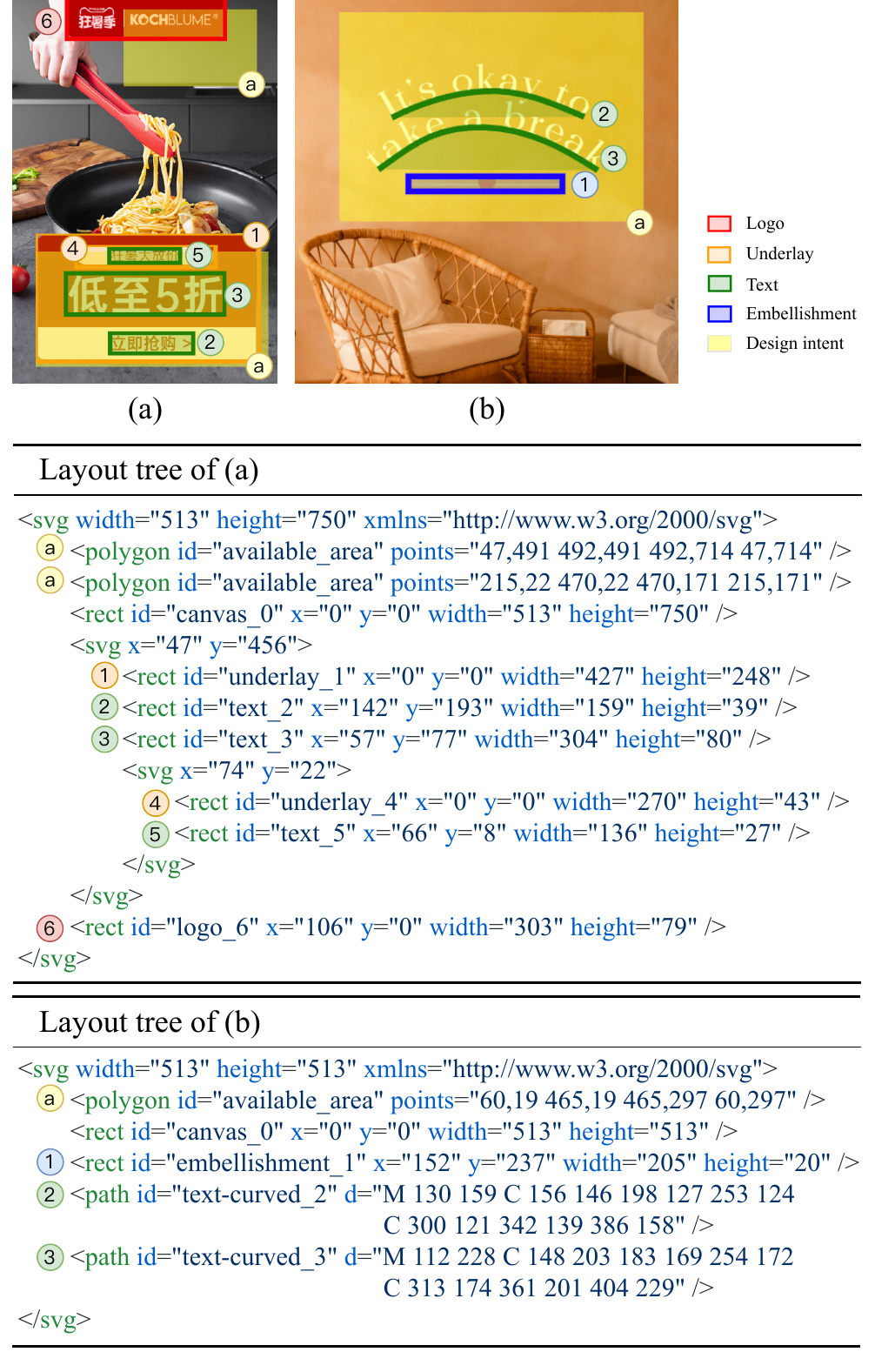}
    \vspace{-2\baselineskip}
    \caption{Examples of layout trees $T$ from (a) PKU PosterLayout and (b) PStylish7 datasets.}
    \label{fig:supp_tree}
    \vspace{-1\baselineskip}
\end{figure}

\paragraph{Layout trees $T$.}
\cref{fig:supp_tree} presents examples of layout trees structured as depicted in \cref{subsec:ltc}.
In (a), two distinct areas \!\textcircled{\small a}\! are detected for its design intent, and the double-nested underlay elements (\textit{i.e.}, \textcircled{\small 1}-\textcircled{\small 5}) contribute to a layout tree depth of three, based on the proposed hierarchical node representation.
In (b), the curved texts (\textit{i.e.}, \textcircled{\small 2},\! \textcircled{\small 3}) are outlined using {\ttfamily <path>}, following universal element vectorization.

\begin{figure}[b]
    \centering
    \vspace{-2.2\baselineskip}
\includegraphics[width=0.96\linewidth]{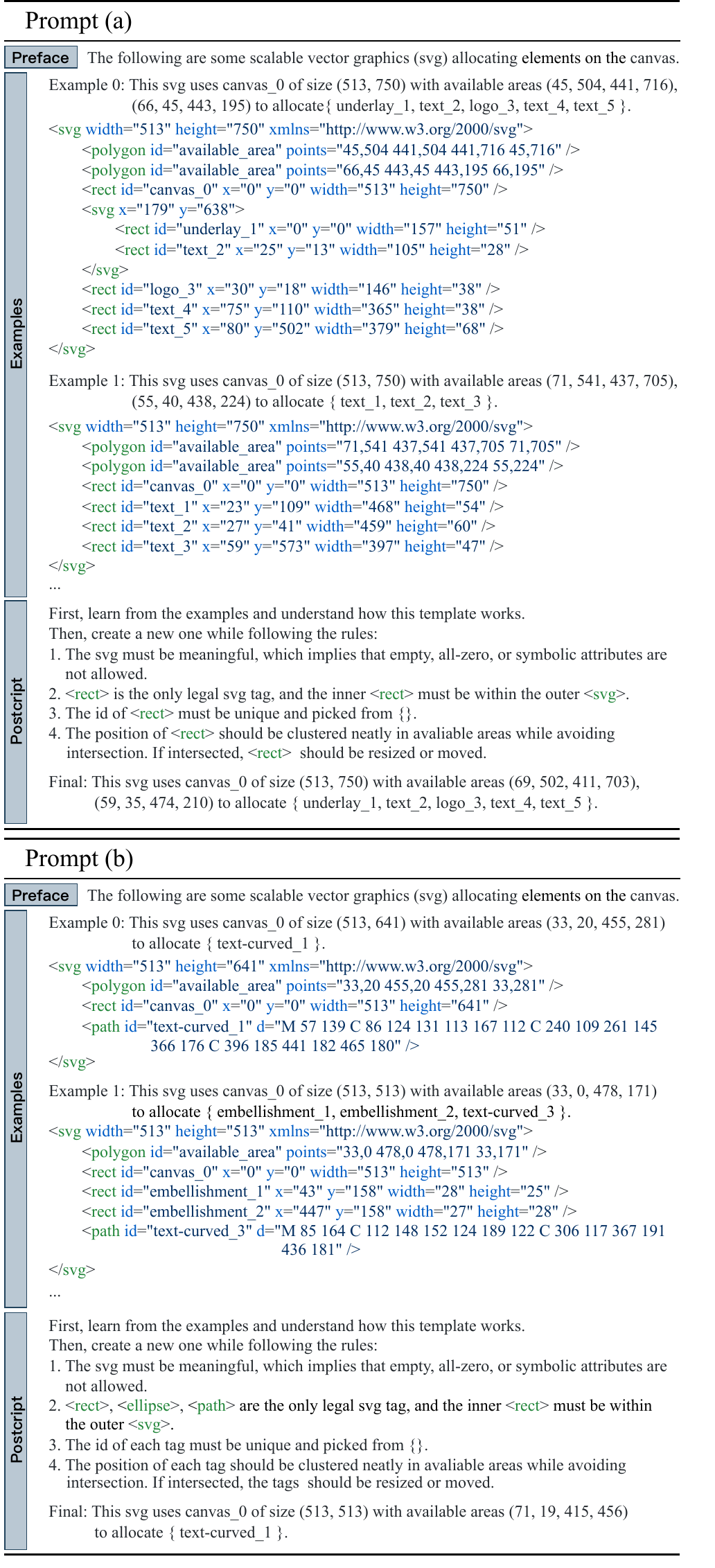}
    \vspace{-0.8\baselineskip}
    \caption{Examples of prompts $P$ on (a) PKU PosterLayout and (b) PStylish7 datasets.}
    \label{fig:supp_prompt}
    \vspace{-1\baselineskip}
\end{figure}

\vspace{-1\baselineskip}
\paragraph{Prompts $P$.}
\cref{fig:supp_prompt} presents examples of prompts created as depicted in \cref{subsec:ltg}.
The canvases' varying sizes demonstrate PosterO's adaptability to diverse aspect ratios.
\begin{figure*}[t]
    \centering
\includegraphics[width=0.88\linewidth]{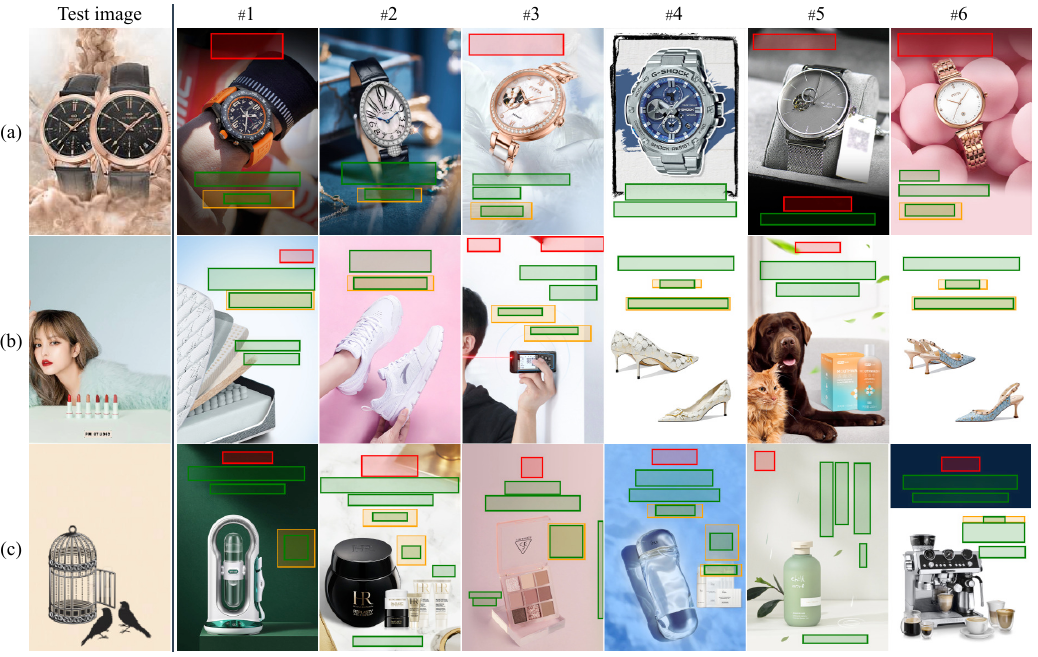}
    \vspace{-0.5\baselineskip}
    \caption{Intent-aligned example selection results ({\small \#}*) on the \textit{unannotated test split} of
PKU PosterLayout dataset.}
    \label{fig:supp_example}
    \vspace{-1\baselineskip}
\end{figure*}

\begin{figure*}
    \centering
    \begin{minipage}{0.98\linewidth}
    \vspace{0.5\baselineskip}
\includegraphics[width=\linewidth]{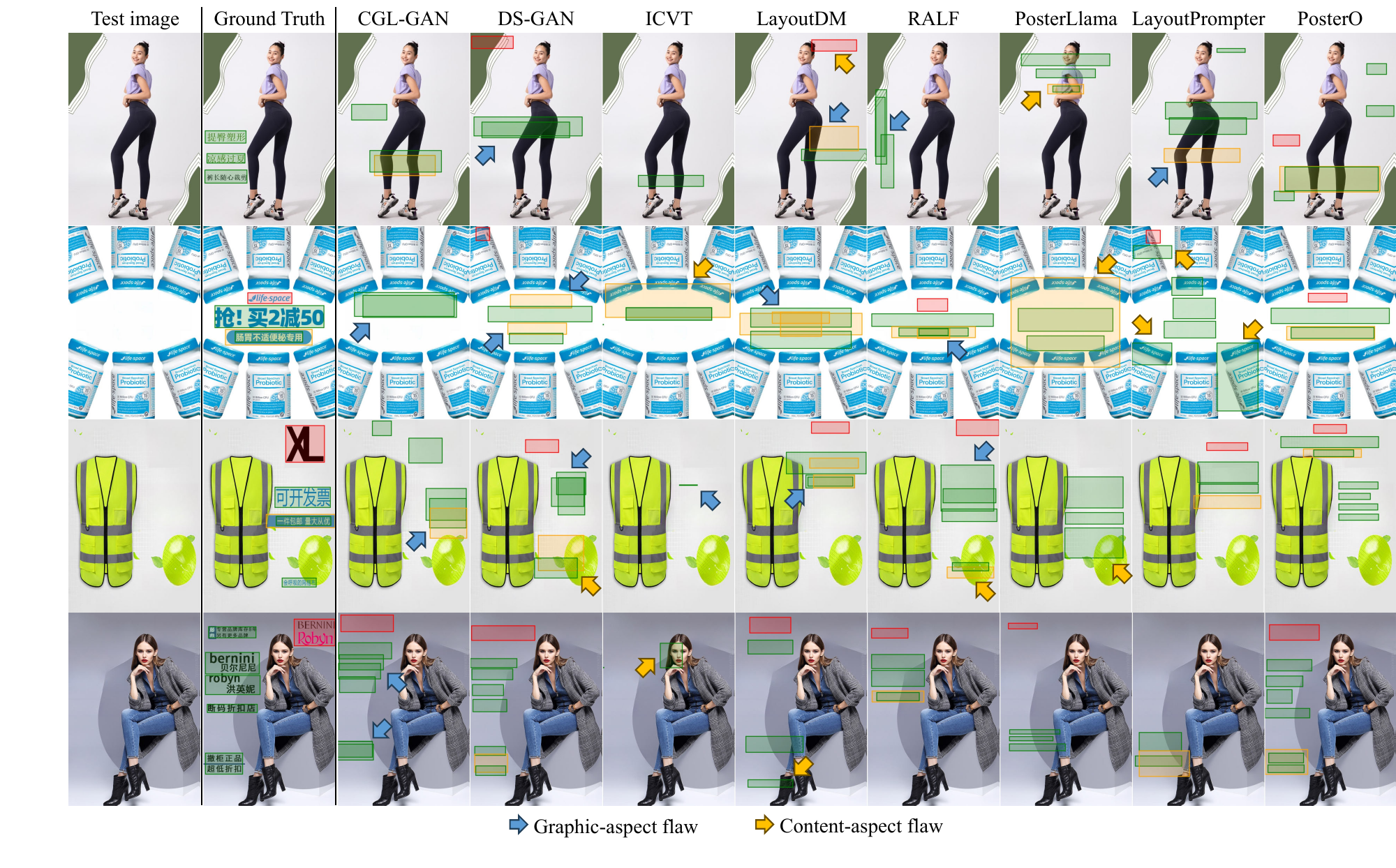}
    \vspace{-1.6\baselineskip}
    \captionof{figure}{Visualized results on the \textit{annotated test split} of
PKU PosterLayout dataset.}
    \vspace{0.4\baselineskip}
    \label{fig:supp_pku_vis}
    \end{minipage}
    \begin{minipage}{0.98\linewidth}
\includegraphics[width=\linewidth]{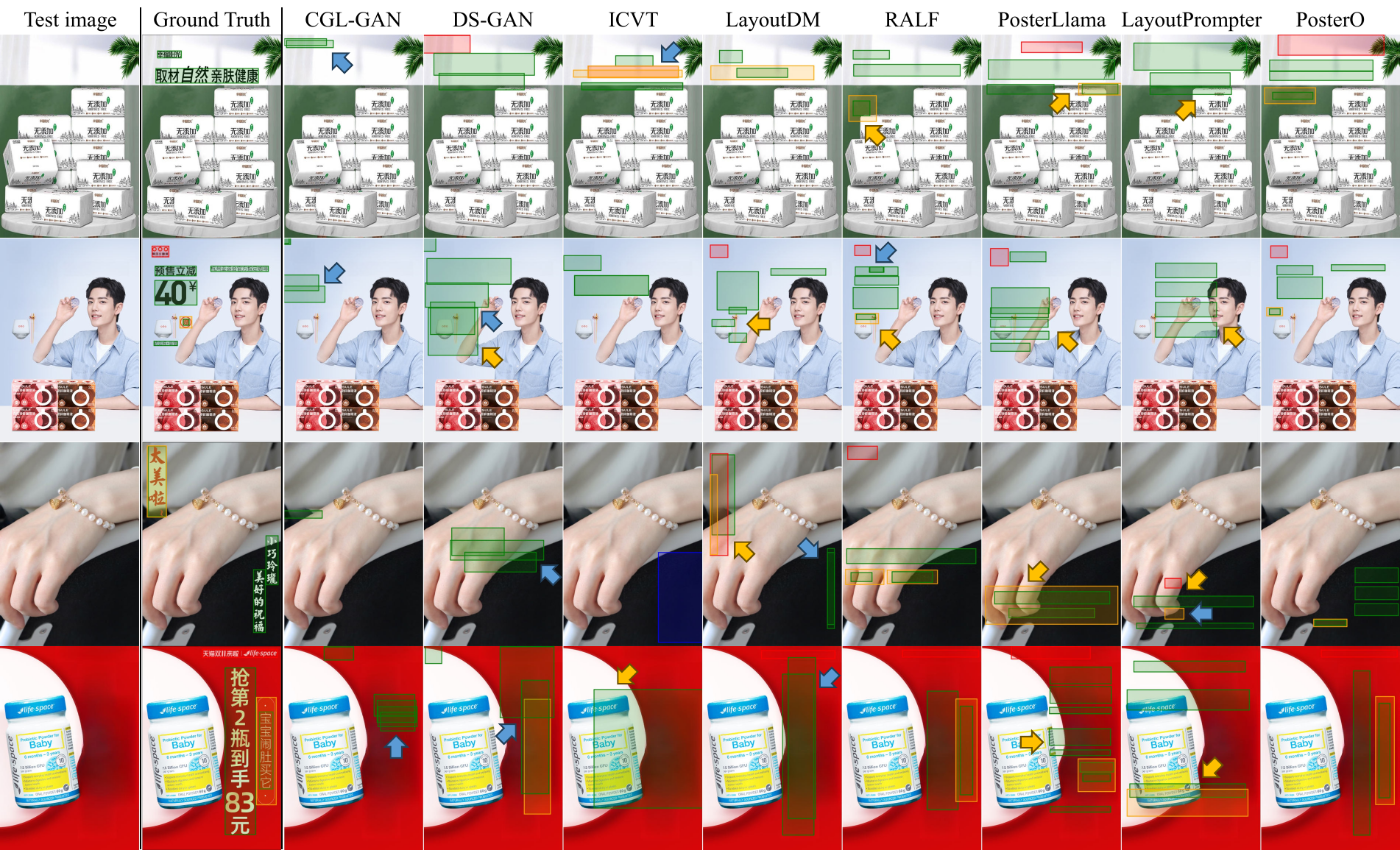}
    \vspace{-1.4\baselineskip}
    \captionof{figure}{Visualized results on the \textit{annotated test split} of
CGL dataset.}
    \label{fig:supp_cgl_vis}
    \vspace{-1\baselineskip}
    \end{minipage}
\end{figure*}

\section{Additional Experimental Results}
\label{sec:supp_result}


\vspace{-0.1\baselineskip}
\paragraph{Selected in-context examples.} 
\cref{fig:supp_example} shows the $k$ layout trees $\{T_j\}^k_j$ selected as in-context examples that align with the intents of the given test images.
By observing (a), we see that the design intent latent space $\textbf{Z}_D$ effectively models both semantics and spatial distribution of objects within the images.
While the composition of the test images appears similar in (b) and (c), the selection results are very different.
These findings demonstrate that our selection procedure provides a comprehensive understanding of visual contents and is sensitive to subtle differences.

\vspace{-1.1\baselineskip}
\paragraph{More visualized results.}
\cref{fig:supp_pku_vis} and \cref{fig:supp_cgl_vis} visualize the layouts generated by all baselines \cite{Min-2022-IJCAI-CGL,hsu-2023-CVPR-posterlayout,Cao-2022-ACMMM-ICVT,horita-2024-CVPR-RALF,inoue-2023-CVPR-layoutdm,lin-2023-NIPS-layoutprompter,seol-2024-ECCV-posterllama} and the proposed PosterO.
Elements violating graphic metrics and content metrics are respectively indicated by blue and yellow arrows.
The comparisons with ground truth demonstrate that PosterO most accurately predicts the design intents and actively utilizes the available areas.
Besides, it understands the graphic relationship in layouts well and generates non-overlapping, aligned elements.

\begin{table}[h]
    \centering
\resizebox{0.995\linewidth}{!}{%
\begin{tabular}{l|cc|ccc|c} \toprule
&
$Und_l\! \uparrow$ &
$Und_s\! \uparrow$ &
$Int\! \downarrow$ &
$Sal\! \downarrow$ &
$Rea\! \downarrow$ &
$Avg\! \downarrow$ \\ \midrule
Ours & \textbf{0.9856} & \textbf{0.9241} & \textbf{0.1427} & \uline{0.2131} & \uline{0.0248} & \textbf{0.0677} \\ \midrule
\rowcolor{Black!20}
\multicolumn{7}{l}{\small Visual condition vectorization (Ours: Design intent)} \\
Saliency & \uline{0.9694} & \uline{0.9114} & 1.2470 & 1.1867 & 0.0374 & 0.3702 \\
\midrule
\rowcolor{Black!20}
\multicolumn{7}{l}{\small Visual information perception (Ours: \textit{d+f})} \\
\textit{-d-f}   & 0.6873 & 0.5776 & 1.1941 & 0.4980 & 0.0275 & 0.3512 \\ 
\textit{+v}     & 0.8784 & 0.6092 & 0.4868 & 0.4145 & 0.0270 & 0.2063 \\
\textit{+v-d}   & 0.8874 & 0.7429 & \uline{0.3142} & \textbf{0.2078} & \textbf{0.0239} & \uline{0.1311} \\
\textit{+v-d-f} & 0.8335 & 0.6399 & 0.6432 & 0.6116 & 0.0302 & 0.2591 \\ \bottomrule
\end{tabular}
}
    \vspace{-0.8em}
    \caption{More ablation studies on visual condition vectorization and visual information perception. (\textit{d}: Design intent vectorization, \textit{f}: Intent-aligned example selection, \textit{v}: LLaMA-3-LLaVA-NeXT.)}
    \label{tab:supp_vision}
    \vspace{-1.4\baselineskip}
\end{table}

\begin{table*}
\centering
\begin{minipage}{0.224\textwidth}
    \includegraphics[width=\textwidth] {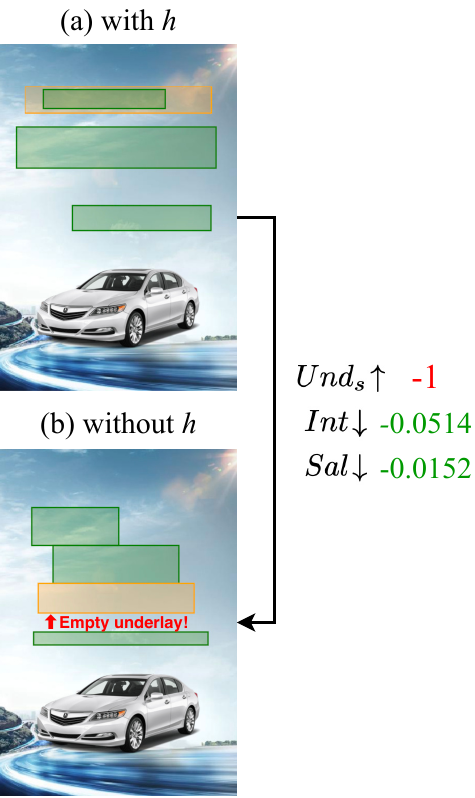}
    \vspace{-1.6em}
    \captionof{figure}{Impact of removing \textit{h}: hierarchical nodes in \cref{tab:ab_layouttree}.}
    \label{fig:supp_hneg}
\end{minipage}
\begin{minipage}{0.6\textwidth}
    \includegraphics[width=\textwidth]{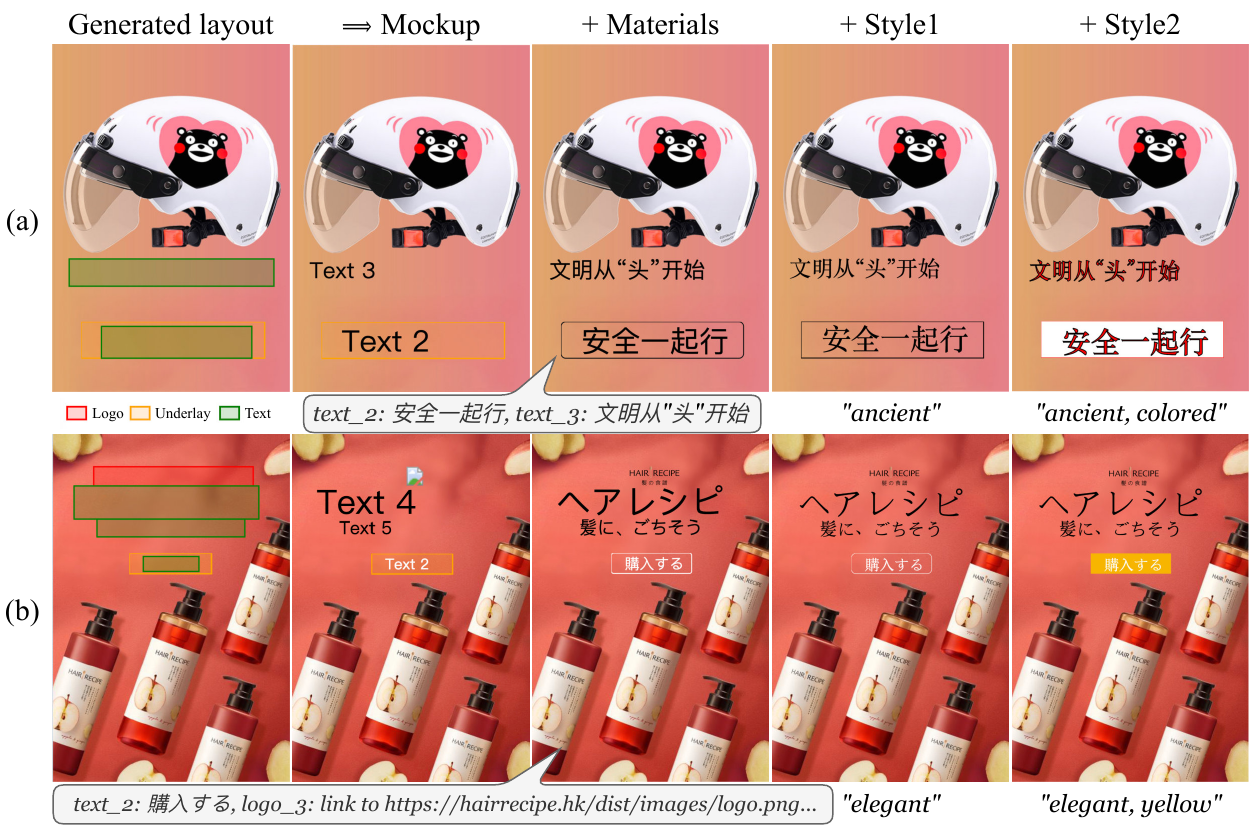}
    \vspace{-1.6em}
    \captionof{figure}{Results of poster design realization.}
    \label{fig:supp_pdr}
\end{minipage}
  \vspace{-1\baselineskip}

\end{table*}

\vspace{-1.1\baselineskip}
\paragraph{More ablation studies on vision processing.}
As reported in \cref{tab:supp_vision}, we investigate more configurations as follows.

(1) Visual condition vectorization:
To demonstrate the advantages of design intent over the widely utilized salient regions, the variant `Saliency' as in LayoutPrompter \cite{lin-2023-NIPS-layoutprompter} is implemented.
It is observed that all content metrics severely drop, surprisingly, even worse than not performing any visual perception (\textit{-d-f}).
This underscores that design intent is an encouraging substitute for the current saliency paradigm.

(2) Visual information perception:
To demonstrate our (\textit{d}, \cref{subsec:ltc}) vectorized conditions and (\textit{f}, \cref{subsec:ltg}) learning examples have already considered input images very well, three variants are implemented with (\textit{v}) LLaMA-3-LLaVA-NeXT\footnote[4]{https://huggingface.co/llava-hf/llama3-llava-next-8b-hf}.
It is a large multimodal model (LMM) with powerful visual reasoning capabilities based on Llama 3-8B.
However, \textit{v} only marginally improves $Sal\!\downarrow$ and $Rea\!\downarrow$ when it replaces \textit{d}, and the rest variants obtain poor results.
As our detection model (16M) is much smaller than LLaVA-NeXT's vision head (encoder: 304M, connector: 20M), \textit{d} is a proper alternative.
In addition, using \textit{d} avoids cross-modal differences between visual tokens and geometric layout elements, reflecting on its better graphic performance.

\begin{table}[t]
\resizebox{\linewidth}{!}{%
\begin{tabular}{r|ccccccc}
\toprule
{\small Ours$_{L2}$} & \textit{Paint} & \textit{Poem}  & \textit{Metro} & \textit{Movie} & \textit{Menu}  & \textit{Animal} & \textit{Insta.} \\
\midrule
$Ove \downarrow$ & 0.0094 & 0.0068 & 0.2622 & 0.0094 & 0.0063 & 0.0085 & 0.0004 \\
$Int \downarrow$ & 0.4159 & 0.2402 & 0.6382 & 0.7213 & 0.1260 & 0.3516 & 0.1408 \\
$Sal \downarrow$ & 3.1897 & 0.6106 & 0.5532 & 0.5806 & 0.4312 & 1.6667 & 0.0590 \\
\midrule
$Avg \downarrow$ & 1.2050 & 0.2859 & 0.4845 & 0.4371 & 0.1879 & 0.6756 & 0.0668 \\
\bottomrule
\end{tabular}%
}
\vspace{-0.5\baselineskip}
\caption{More quantitative results on PStylish7 dataset.}
\vspace{-1.3\baselineskip}
\label{tab:supp_pstylish}
\end{table}

\vspace{-1.1\baselineskip}
\paragraph{`Negative' impact of hierarchical nodes.}
As the enclosing structure is equivalent to \textit{nested layouts}, the current design of (\textit{h}, \cref{subsec:hier}) hierarchical node representation has effectively constrained the solution space.
However, by observing the first two rows of \cref{tab:ab_layouttree}, removing \textit{h} appears to improve the standardized content metrics, $Int\!\downarrow$ and $Sal\!\downarrow$.
To determine the primary cause, we analyzed the visualization results and found a significant occurrence of \textit{empty underlays}, as depicted in \cref{fig:supp_hneg}.
These invalids improve in $Cov$ and $Uti$, thereby deceptively enhancing the standardized metrics.
This finding again highlights the importance of \textit{h} in generating layouts with satisfactory integrity.

\vspace{-1.1\baselineskip}
\paragraph{Results of poster design realization.}
\cref{subsec:pdr} introduces the zero-shot transformation from generated layouts to actual posters.
\cref{fig:supp_pdr} shows the results of each step and the final designs.
During \textit{mockup creation}, not only the font size and text color are properly predicted, but also the link to the logo image, \textit{i.e.}, {\ttfamily href} attribute, is correctly created with an initial unknown value.
For \textit{material synthesis}, all given contents are perfectly placed into the elements of corresponding {\ttfamily id}.
We also request LLMs with \textit{style} in simple words, such as \textit{elegant}, and the resultant posters are astonishingly appealing.

\begin{figure}[t]
    \centering
\includegraphics[width=0.95\linewidth]{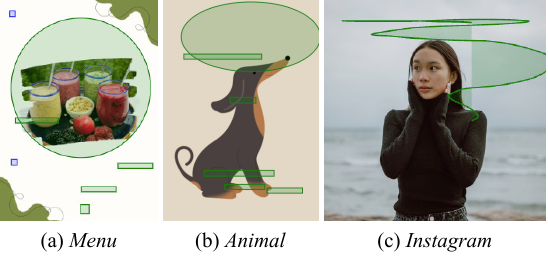}
    \vspace{-0.3\baselineskip}
    \caption{Visualized results of PosterO$_{L2}$ on PStylish7 dataset.}
    \label{fig:supp_ps}
    \vspace{-1.2\baselineskip}
\end{figure}

\vspace{-1.1\baselineskip}
\paragraph{PosterO$_{L2}$ on PStylish7.}
In the main text, \cref{tab:pstylish} reported the quantitative results of PosterO$_{CL}$ and PosterO$_{L3}$ on PStylish7.
Here, \cref{tab:supp_pstylish} reports those of PosterO$_{L2}$.
Surprisingly, it achieves the best overall performance on \textit{Metro}, \textit{Menu}, \textit{Animal}, and especially \textit{Instagram}.
By accumulating $Avg\!\downarrow$ across seven categories, PosterO$_{L2}$, $_{CL}$, and $_{L3}$ obtain \textbf{3.3427}, 4.3799, and \uline{4.2877}, respectively.
We visualize some amazing layouts generated by PosterO$_{L2}$ in \cref{fig:supp_ps}.

\section{PStylish7: More Details and Statistics}
\label{sec:supp_ps7}

PStylish7 is the first dataset for generalized content-aware layout generation.
We built it with 152 image-layout pairs for few-shot learning and 100 image canvases for benchmark testing.
The data comes from a variety of sources, including Canva, Pixabay, Ucshe, and the New York Heritage Digital Collections.
In the remaining section, statistics on the distribution of sample categories, element types, and image aspect ratios within the PStylish7 dataset are provided to offer a clear understanding of its diversity and complexity.
Additionally, a detailed explanation of the metrics calculation is presented.

\begin{figure*}[t]
    \centering
    \begin{tabular}{ccc}
        \includegraphics[width=0.3\linewidth]{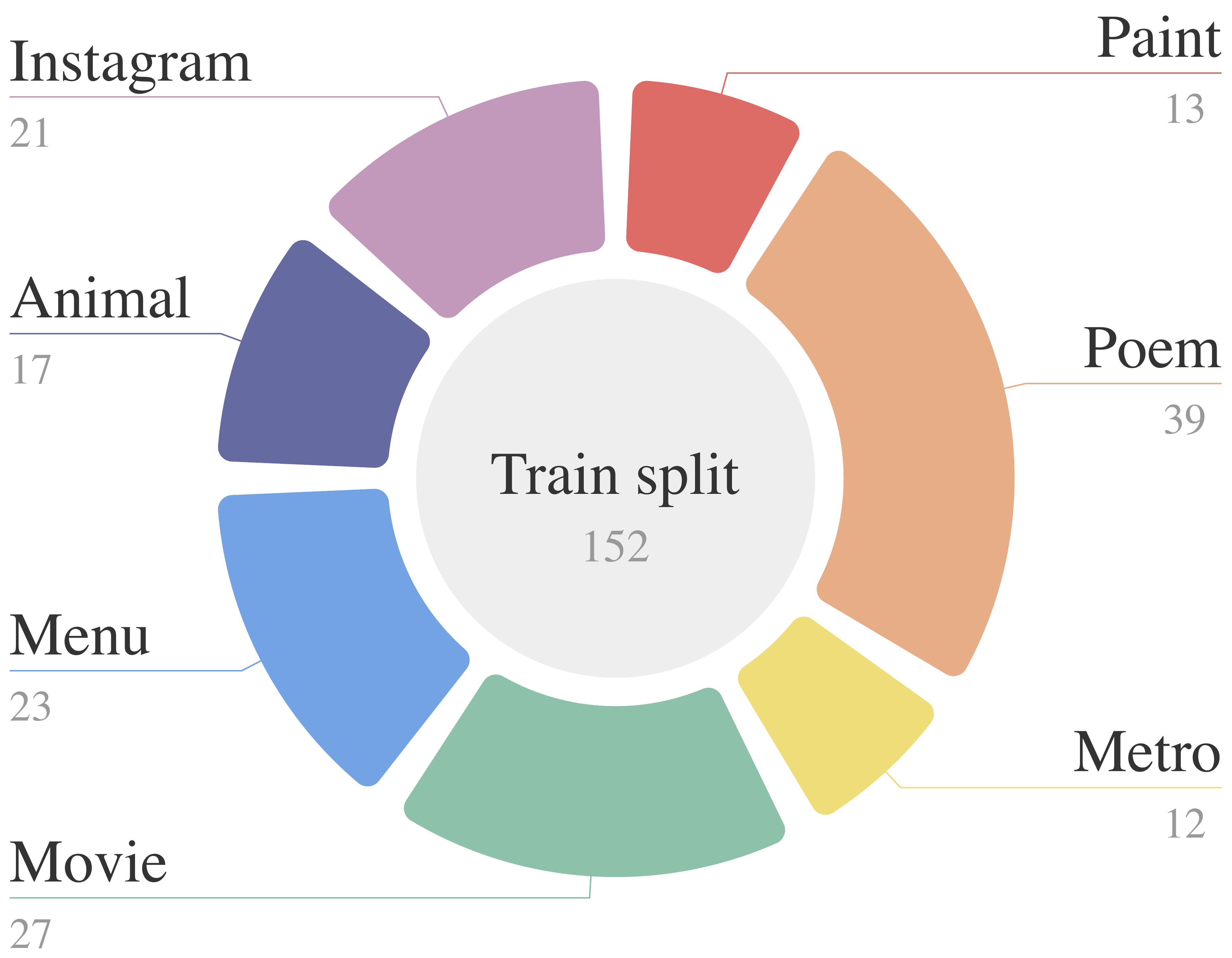} & 
        \includegraphics[width=0.3\linewidth]{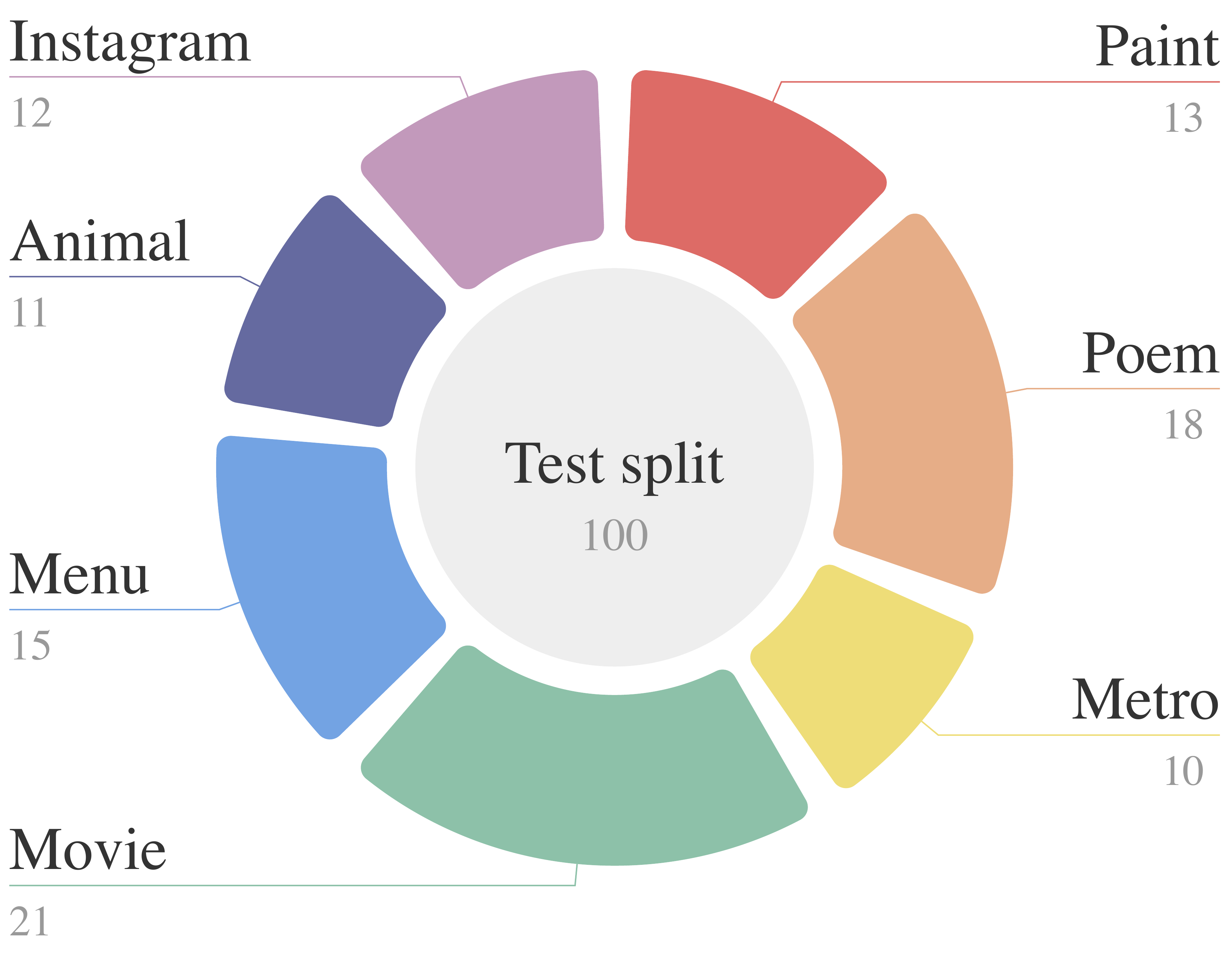} &
        \includegraphics[width=0.3\linewidth]{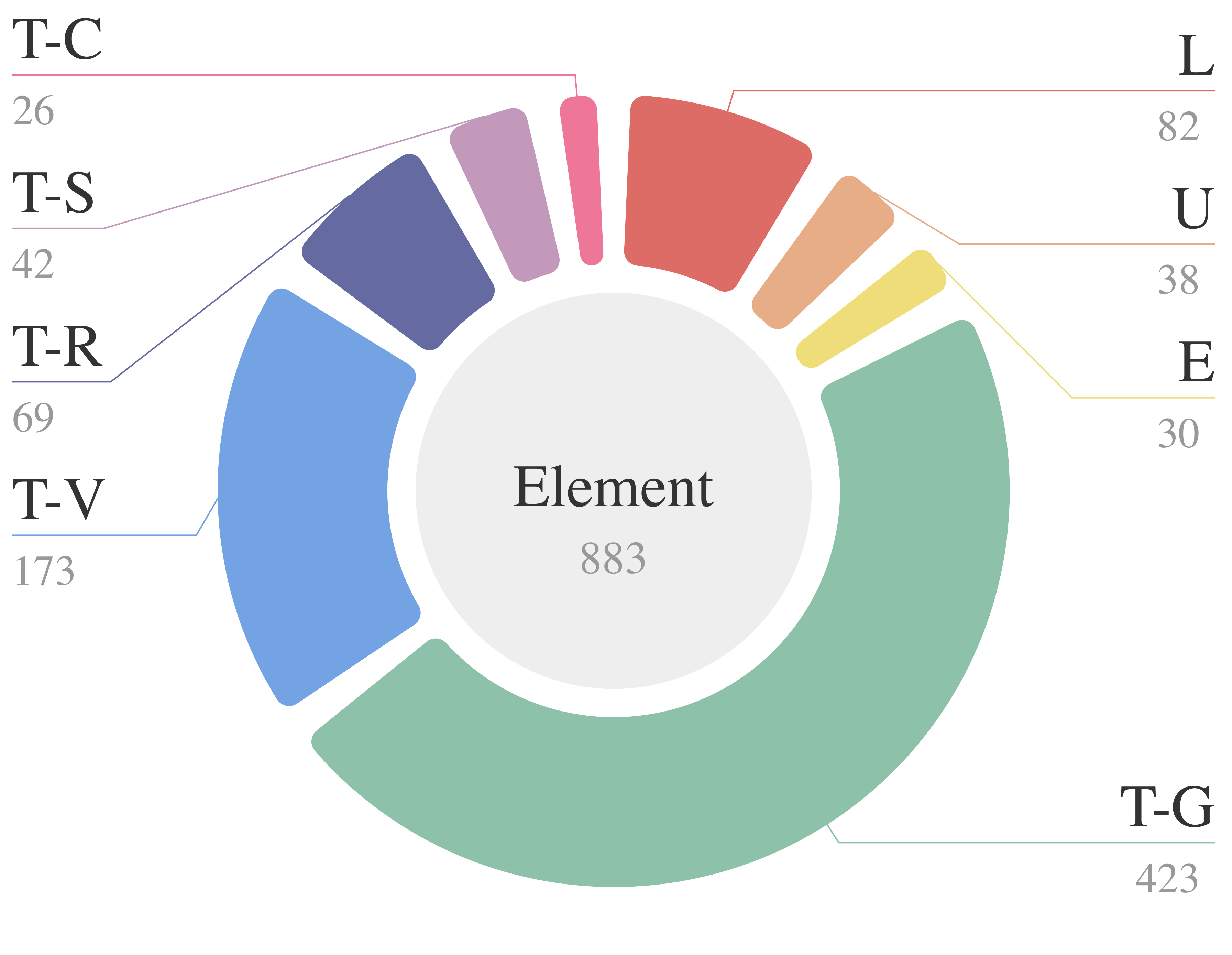} \\
        \small (a) Distribution of poster categories &
        \small (b) Distribution of canvas categories &
        \small (c) Distribution of layout element types
    \end{tabular}
    \vspace{-0.4\baselineskip}
    \caption{Statistics on sample categories and layout element types in PStylish7 dataset.}
    \label{fig:ps_sta}
    \vspace{-1\baselineskip}
\end{figure*}

\subsection{Statistics}

\vspace{-0.1\baselineskip}
\paragraph{Sample categories.}
The seven categories in PStylish7 are artwork exhibition (\textit{Paint}), cultural education (\textit{Poem}), public safety (\textit{Metro}), entertainment marketing (\textit{Movie}), merchandising display (\textit{Menu}), public advocacy (\textit{Animal}), and social-media interaction (\textit{Instagram}).
The distributions of samples across different categories in the train and test splits are illustrated in \cref{fig:ps_sta}(a) and (b).

\vspace{-1.1\baselineskip}
\paragraph{Inter-category gaps.}
\begin{figure}[t]
    \centering
\includegraphics[width=\linewidth]{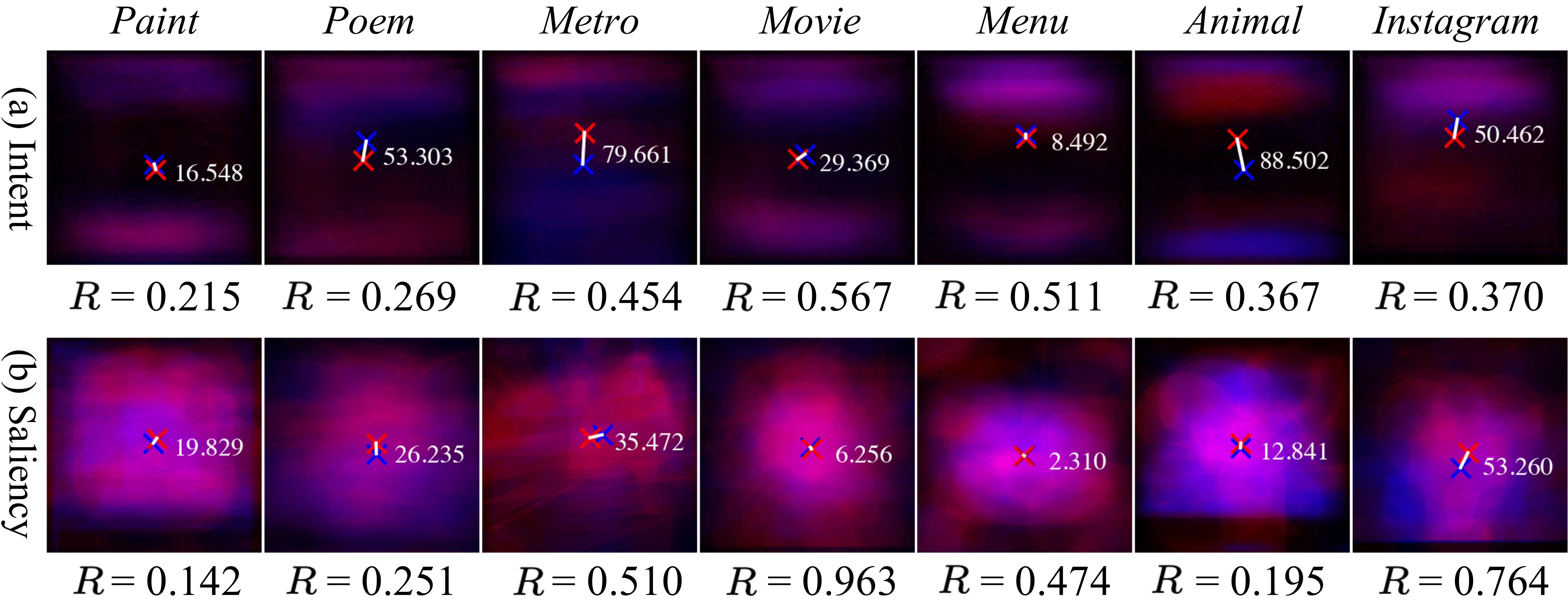}
    \vspace{-1.2\baselineskip}
    \caption{Varied difficulty degrees of PStylish7 categories.}
    \label{fig:ps_gaps}
    \vspace{-1.3\baselineskip}
\end{figure}
With the diverse purposes and entities in different categories, their difficulty degrees $H$ are also noticeably varying.
To gain insight into $H$, we analyzed each category by visualizing the distribution variation $\Delta D$ between their {\color{blue} train}/{\color{red} test} splits and indicating the trade-off factors $R\!=\!\frac{\textit{Unmatch}_L}{\textit{Match}_L}$ for content metrics, as shown in \cref{fig:ps_gaps}.
As observed, $H\! \propto\! \frac{\Delta D}{R}$, aligning with PosterO's performance reported in \cref{tab:pstylish} and \cref{tab:supp_pstylish}.

\vspace{-1.1\baselineskip}
\paragraph{Layout element types.}
The eight layout element types within PStylish7 are logo (L), underlay (U), embellishment (E), and text (T-G, general), along with four text variants (T-V, vertical; T-R, rotated; T-S, ellipse; T-C, complex curve), which are exclusive to this new dataset.
The distribution of elements across different types is illustrated in \cref{fig:ps_sta}(c), showing a total of 883 elements.
Notably, over one-third of them are specialized variants, highlighting the dataset's emphasis on capturing the complexity inherent in real-world design work.

\begin{figure}[t]
    \centering
\includegraphics[width=\linewidth]{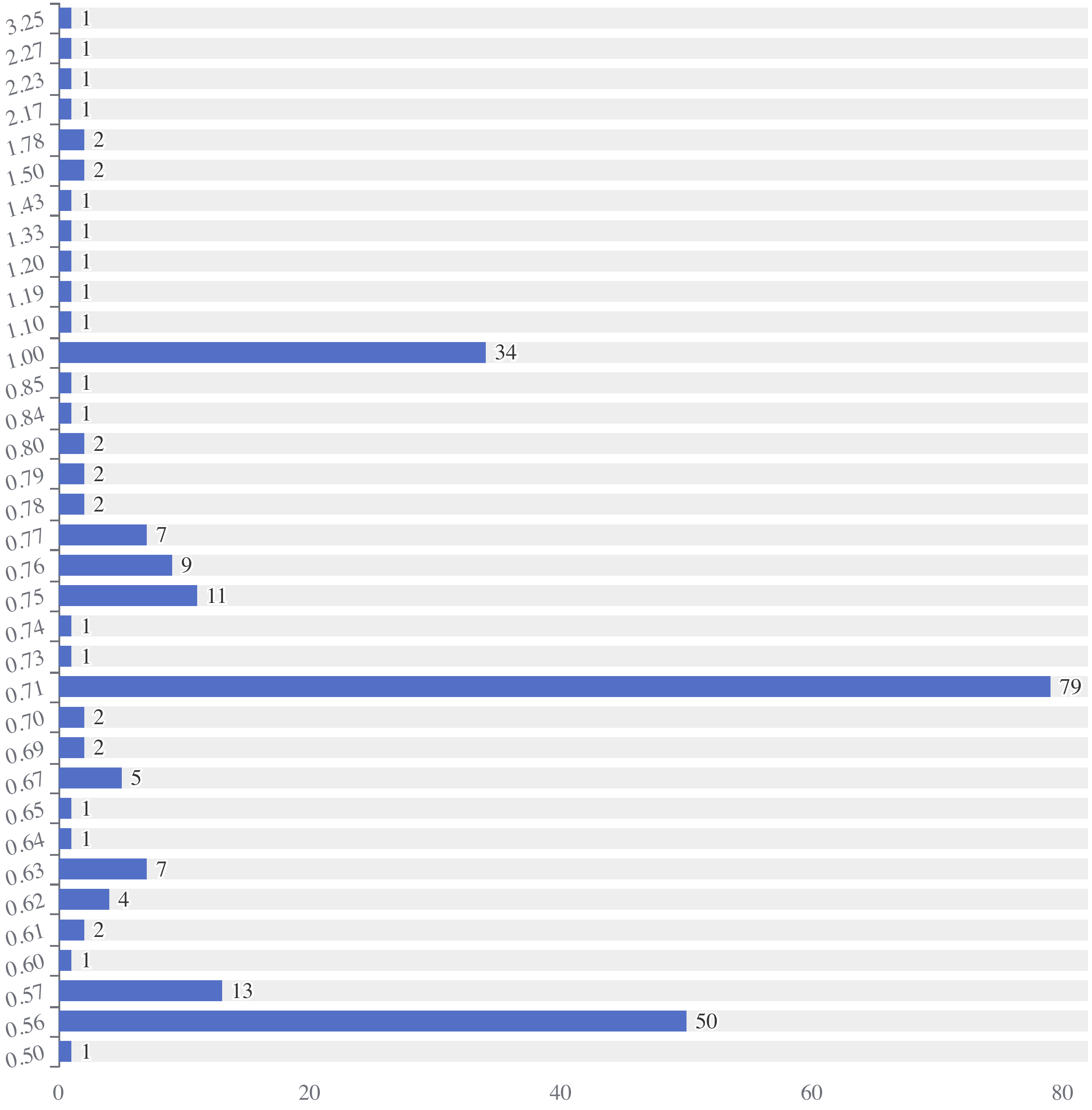}
    \vspace{-1.5\baselineskip}
    \caption{Distribution of image aspect ratios in PStylish7 dataset.}
    \label{fig:ps_aspect}
    \vspace{-1.2\baselineskip}
\end{figure}

\vspace{-1.1\baselineskip}
\paragraph{Image aspect ratios.}
The distribution of image aspect ratios within PStylish7 is illustrated in \cref{fig:ps_aspect}.
Unlike existing domain-specific datasets \cite{hsu-2023-CVPR-posterlayout,Min-2022-IJCAI-CGL}, where samples are predominantly of a fixed aspect ratio (\textit{i.e.}, 0.68), PStylish7 includes a variety of aspect ratios, with the most common being 5:7 (\textit{i.e.}, 0.71), 9:16 (\textit{i.e.}, 0.56), and 1:1.
This highlights the dataset's emphasis on providing a more realistic conditions for layout generation tasks. 

\subsection{Metrics}
Given the high flexibility of the specialized elements, current numerical metrics (\textit{e.g.}, $Ali\!\downarrow$) often fail to evaluate their placement and organization.
To this end, we resort to pixel-level metrics and regularize the rendering process of element maps for computation.
While the content metrics (\textit{i.e.}, ($Cov\!\uparrow$, $Con\!\downarrow$), standardized as $Int\!\downarrow$, and ($Uti\!\uparrow$, $Occ\!\downarrow$), standardized as $Sal\!\downarrow$) are inherently based on element maps, the graphic ones are not.
Therefore, we introduce a pixel-level overlay $Ove\!\downarrow$ to fill in this vacancy.

\vspace{-1.1\baselineskip}
\paragraph{Element map rendering process.}
An element map $m_{e_i} \!\in\! \{0, 1\}^{h\times w}$ serves as the visual indicator revealing the spatial coverage of layout elements $\{e_i\}_i^n$.
\cref{fig:ps_element_map} illustrates the corresponding element maps of layouts in \cref{fig:supp_ps}.
While rectangular elements are straightforwardly filled with white, those approximated by ellipses and curves require a different way.
Concretely, we utilize strokes with a width of 30 pixels to outline these elements.
Every element map is uniformly scaled to a width $h$ of 513 pixels to maintain consistency in scale for appropriate comparisons.
With our SVG language-based representations, the rendering can be easily implemented by introducing CSS-style declarations as:
\begin{Verbatim}[fontsize=\small]
rect { fill: white; }
ellipse { stroke: white; stroke-width: 30; }
path { stroke: white; stroke-width: 30; }
\end{Verbatim}

\begin{figure}[t]
    \centering
\includegraphics[width=0.95\linewidth]{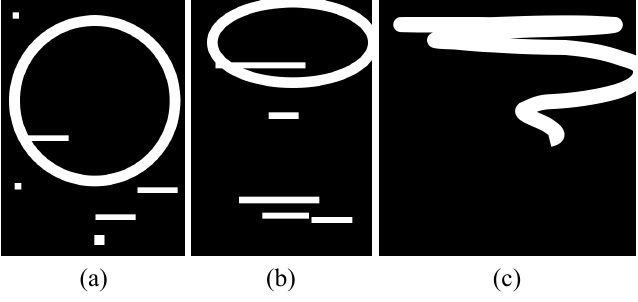}
    \vspace{-0.5\baselineskip}
    \caption{Rendered element maps of layouts in \cref{fig:supp_ps}.}
    \label{fig:ps_element_map}
    \vspace{-1.4\baselineskip}
\end{figure}

\vspace{-1.1\baselineskip}
\paragraph{Pixel-level overlay $Ove\!\downarrow$.}
To calculate overlay between non-underlay elements $\{e_j\}_j^m$, each element $e_j$ is rendered individually as a layer $\ell_j$ of the element map.
The overlay is then determined by calculating the sum of layers $\{\ell_j\}_j^m$ and subtracting the map $m_{e_j}$ that results from their combination.
Specifically, as the maximum value of each pixel in the map is 1, this pixel-level operation accumulates the number of pixels where more than one element is present.
By dividing the size of the map $m_{e_j}$, we obtain a normalized measure of overlay.



\end{document}